\documentclass{aa} 

\usepackage{graphicx}
\usepackage{subcaption}
\usepackage{upgreek}
\usepackage{hyperref}
\usepackage{txfonts}
\usepackage[separate-uncertainty=true]{siunitx}
\DeclareSIUnit\erg{erg}
\DeclareSIUnit\parsec{pc}
\DeclareSIUnit\years{yrs}

\begin{document} 

   \title{ViCTORIA project: The LOFAR-MeerKAT view of AGN in Virgo cluster early-type galaxies}

   \subtitle{}

   \author{A.~Spasic
          \inst{1},
          H.~W.~Edler\inst{1},
          Y.~Su\inst{3},
          M.~Brüggen\inst{1},
          F.~de~Gasperin\inst{2},
          T.~Pasini \inst{2},
          V.~Heesen \inst{1}
          M.~Simonte \inst{1}
          A.~Boselli \inst{4}, 
          H.~J.~A.~Röttgering \inst{5},
          M.~Fossati \inst{6,7}
          }

   \institute{Hamburger Sternwarte, University of Hamburg, Gojenbergsweg 112, D-21029, Hamburg, Germany
         \and
             INAF - Istituto di Radioastronomia, via P. Gobetti 101, Bologna, Italy
         \and
             Department of Physics and Astronomy, University of Kentucky, 505 Rose Street, Lexington, KY 40506, USA
         \and
             Aix Marseille Univ., CNRS, CNES, LAM, 13388 Marseille Cedex 13, France
         \and 
             Leiden Observatory, Leiden University, PO Box 9513, NL-2300 RA Leiden, the Netherlands
        \and 
            Università degli studi di Milano-Bicocca, Piazza della scienza 3, Milano, Italy
        \and 
            INAF - Osservatorio Astronomico di Brera, via Brera 28, Milano, Italy
             }

   \date{Received Month date, 202x; accepted Month date, 202x}

 
  \abstract
   {The evolution of Active Galactic Nuclei (AGN) is closely connected to their host galaxies and surroundings. Via feedback processes, AGN can counteract the cooling of the intracluster medium (ICM) and suppress star formation in their host galaxies. Radio observations at low frequencies provide a glimpse into the history of AGN activity. The Virgo cluster is a substantial reservoir of nearby galaxies and provides an ideal laboratory for the study of AGN as well as their feedback mechanisms.}
   {The aim of our work is to characterise the AGN population within the Virgo cluster down to low radio luminosities, constrain the AGN duty cycle and investigate environmental feedback in cluster member galaxies.}
   {We analyse 144\,MHz and 1.3\,GHz radio observations of early-type galaxies from the ACS Virgo Cluster Survey (ACSVCS) taken with LOFAR and MeerKAT.}
   {We detect 12 of these galaxies at 144\,MHz, 5 of which show clearly extended radio emission. The radio luminosity shows a strong dependence on the stellar mass of the host galaxy, in agreement with previous results. As a notable outlier, the massive elliptical galaxy NGC\,4365 ($M_*=2.2\times10^{11}\,M_\odot$) is not detected as compact source in the LOFAR observations. Instead, it is surrounded by diffuse, low-surface brightness emission, which hints towards a past phase of stronger nuclear activity. Furthermore, we find a cavity in NGC\,4472 (= M\,49) inflated by the wide-angle tail only visible in the LOFAR data, which implies that the cavity was created by a past outburst. The corresponding cavity power is of the same order of magnitude as the jet power in the present duty cycle of the AGN.}
   {}


    \authorrunning{Spasic, A., Edler, H. W., et al.}
   \maketitle
%

\section{Introduction}
Most nearby galaxies harbour a supermassive black hole (SMBH) in their centre. By accreting matter (mostly gas and dust) via an accretion disk forming around the SMBH, the galaxy can manifest an active galactic nucleus (AGN, \citealt{1963MNRAS.125..169H}). These AGN can produce emission over a wide range of the electromagnetic spectrum and are complicated objects in terms of their phenomenology. Most characteristically, they can produce two collimated jets of relativistic particles.

A primary way to probe these jets is the observation of their non-thermal synchrotron emission in radio galaxies. As AGN show a large variety of radio properties and morphologies, the study of radio emission gives important insights on the mechanisms feeding the nuclear activity. Many scaling relations between the radio emission from AGN, their properties and the host galaxy properties were identified in the past \citep[e. g.][]{2005MNRAS.362...25B, 2013ARA&A..51..511K, 2022A&A...660A..93C}.

Analysing a large sample of radio-loud galaxies with $z \leq 0.3$, \cite{2005MNRAS.362...25B} find a strong correlation between the radio-loud AGN fraction and host mass. This correlation has been confirmed in later studies, as in \cite{2019A&A...622A..17S} using LOFAR data, concluding that all galaxies with $M > 10^{11}\,\mathrm{M_\odot}$ host a radio-loud AGN. Similar results were found by \cite{2022ApJS..258...30G} in a sample of nearby, X-ray-bright galaxies, and by \cite{2022A&A...660A..93C} in their study of giant early-type galaxies in the nearby Universe. They also find a connection between the size of the radio source and its luminosity, arguing that the jet properties of point-sources differ from those of extended sources.

The morphology, extent and the spectral properties of the extended radio emission allow us to probe the evolution of the jet activity. Since relativistic particles emitting at the highest radio frequencies lose energy faster than those emitting at lower frequencies, the spectrum of synchrotron emission steepens with time \citep{1962SvA.....6..317K,1973A&A....26..423J,2013MNRAS.435.3353H}. This spectral steepening causes AGN plasma to be only observable only for a relatively short time at Gigahertz-frequencies and above. Low-frequency observations (<\SI{300}{\mega\hertz}), such as those conducted by the LOw-Frequency ARray \citep[LOFAR,][]{2013A&A...556A...2V}, are therefore especially important to trace the oldest outbursts and the past evolution of AGN.

In order to study AGN-related radio emission, it is important to differentiate from the radio emission produced by star formation. As this can become very complicated for star-forming galaxies, a lot of studies focus on AGN in early-type galaxies, where the contribution to radio emission from star formation is much smaller \citep[e.g.][]{2005MNRAS.362...25B,2009AJ....138.1990C,2022A&A...660A..93C,2019A&A...622A..17S}. This is also the case in our study.

Jets of radio galaxies can transfer energy and momentum into the surrounding medium, causing a heating of the gas. This re-deposition of energy is known as AGN feedback \citep{2012ARA&A..50..455F}. The process is not only able to quench star formation but also prevents the growth of galaxies via gas accretion \citep{2013ARA&A..51..511K}.

AGN are therefore believed to play an important role in the evolution of their host galaxies \citep{2000ApJ...539L...9F,2006AJ....131...84V}. This can be seen best in the tight correlation of the black hole mass with the masses of elliptical galaxies and bulges \citep{2013ARA&A..51..511K}. On larger scales AGN also play a role in the evolution of the clusters they reside in \citep[e.g.][]{2012NJPh...14e5023M} and are also affected by it \cite{2023ApJ...956..104H}. Multiple studies show that radio galaxies are more common in galaxy-rich environments \cite{2004ApJ...607..800B, 2013MNRAS.430..638S, 2015A&A...576A.101M}. There are, however, still many open questions in this field.

Especially in galaxy clusters, AGN feedback has a large effect on the intracluster medium (ICM). In the so-called radio-mode feedback the jets expel radio-heated gas into the ICM, which pushes the X-ray heated gas away \citep{2020MNRAS.496.1706S}, forming large radio lobes. This creates 'cavities' in the ICM, areas where a lack of X-ray emission can be seen. With these cavities it is possible to estimate the power of the jet that produced them and the energy that is supplied to the surrounding medium \citep{2004ApJ...607..800B,2006ApJ...652..216R, 2008ApJ...686..859B}.

As the nearest massive ($M_\mathrm{vir}=1.0-1.4{\times}10^{14}\,\mathrm{M_\odot}$, \citealt{Urban2011X-rayRadius,Ferrarese2012,2017MNRAS.469.1476S}) galaxy cluster, the Virgo cluster is not only the largest accumulation of early-type galaxies in the local universe, but also a prime target to study AGN feedback. Studies of the system allow us to probe AGN across a vast range of luminosities and host galaxy masses. A study of the early-type galaxies in the Virgo cluster has been done by \citet{2009AJ....138.1990C} using VLA at a frequency of 8.4\,GHz, where they investigated the origin of radio emission and the connection with the properties of the host galaxies. Analysing a sample of the 63 optically brightest ACSVCS galaxies, out of which 12 show the presence of a compact radio source, they find a strong relation between the AGN fraction and stellar mass in agreement with other results. They also find that some massive galaxies show no signs of nuclear activity despite the high mass of the SMBH.

In this paper, we will extend the analysis of the early-type galaxies in Virgo as done by \cite{2009AJ....138.1990C} to lower frequencies, using the high-sensitivity LOFAR and MeerKAT data of the `Virgo cluster multi-telescope observations in radio of interacting galaxies and AGN' (ViCTORIA) project \citep[ de Gasperin et al. in prep.]{2023A&A...676A..24E}. This allows us to study the presence of extended jets and past phases of nuclear activity.  

The structure of the paper is as follows: In Section \ref{sec:data} we describe the sample selection and give an overview of the radio data. Section~\ref{sec:full sample} discusses the radio analysis of the full data sample and analyses the relations between the different properties. In Section~\ref{sec:extended sources} we focus on the properties of the extended sources. A summary and conclusion is given in Section \ref{sec:conclusion}. Throughout this paper we adopt a flat $\Lambda$CDM  cosmology with $H_0 = \SI{70}{\km\per\s\per\mega\parsec}$, $\Omega_M = 0.3$ and $\Omega_{\Lambda} = 0.7$. At the distance of NGC\,4486 (= M\,87) $d = \SI{16.5}{\mega\parsec}$ \citep{2007ApJ...655..144M,2018ApJ...856..126C}, one arcsecond corresponds to $\SI{80}{\parsec}$.
\section{Data and sample}
\label{sec:data}

In our study, we analyse multi-frequency radio data of a sample of early-type galaxies in the Virgo cluster. The radio data were collected in the context of the ongoing ``Virgo Cluster multi-Telescope Observations in Radio of Interacting Galaxies and AGN'' (ViCTORIA) project (de Gasperin et al. in prep.). The final ViCTORIA data set will consists of three blind radio surveys of the cluster, using the LOFAR low-band antenna (LBA, 42--66\,MHz), the LOFAR high-band antenna (HBA, 120--168\,MHz) and MeerKAT (856--1712\,MHz). 
For this study, we will employ the LOFAR HBA data which was published previously in \citet{2023A&A...676A..24E} and is available in the public domain\footnote{\url{https://lofar-surveys.org/virgo_data.html}}. In addition, we carried out the data reduction and analysis of the MeerKAT observations covering the sources in our sample.
In the following section, we will introduce this sample and the data reduction procedure of the radio observations.

\subsection{Sample selection}

Our sample is composed of the galaxies from the ACS Virgo Cluster Survey (ACSVCS), originally defined by \cite{2004ApJS..153..223C}. This survey is an imaging survey with the Advanced Camera for Surveys (ACS) onboard the Hubble Space Telescope on 100 early-type galaxies in the Virgo cluster. The ACSVCS includes only galaxies classified as members of the Virgo cluster according to the Virgo Cluster Catalog (VCC) by \cite{1987AJ.....94..251B} with a known radial velocity. All galaxies were chosen to have a total magnitude of $B_T \geq 16\,$mag and are classified as early-types according to \cite{1984AJ.....89...64B}. Excluding galaxies with strong dust lanes, lacking visible bulge components or showing tidal interactions leaves a subset of 100 galaxies. Many properties of those galaxies have been derived in the context of the ACSVCS. These include the total stellar mass \citep{2008ApJ...681..197P} and a distance calibration from surface brightness fluctuations \citep[recalculated by][]{2009ApJ...694..556B}. We have adopted these parameters for the analyses of this paper. 

We cross-matched the ACSVCS galaxies with the ViCTORIA LOFAR HBA maps discussed in the following Section \ref{subsec:lofar}. This yielded a working sample of 12 early-type galaxies in the Virgo cluster which are detected by LOFAR. This sample overlaps in ten cases with the radio detected sample of \cite{2009AJ....138.1990C}. We find two additional sources (NGC\,4262 and NGC\,4621) which are not radio detected in that work. For two other sources (NGC\,4762 and NGC\,4550) with compact radio emission found by \cite{2009AJ....138.1990C}, we find no significant radio emission with LOFAR.
For the analysis of the extended tails of NGC\,4472 (=\,M\,49), we also re-analyse XMM-Newton observations originally published by \citet{2019AJ....158....6S}.

\begin{table*}
\caption{Image properties and flux density measurements of the sample}            
\label{tab:image_prop}
\setlength{\tabcolsep}{12pt}
\centering        
{\tiny
\begin{tabular}{cccccccc}       
\hline\\[-1.7ex]            
NGC & VCC & LOFAR & $\sigma_\mathrm{rms,144MHz}$& $S_\mathrm{144MHz}$ & MeerKAT & $\sigma_\mathrm{rms,1.3GHz}$ & $S_\mathrm{1.3GHz}$ \\
 &  & pointings  &  [$\mathrm{\mu Jy\,beam^{-1}}$] & [Jy] & pointings &  [$\mathrm{\mu Jy\,beam^{-1}}$] & [Jy]\\
\hline\\[-1.7ex]                      
   4262     &  355 & 2,3  & 240 & $(1.2 \pm 0.3)\times 10^{-3}$ & 93,115 &  &\\      
   4365$^*$ &  731 & 4,7,8 & 277 & $<\num{1.0e-3}$ & 102,123,124,135 & 12.3 & \num{1.9e-4}\\
   4374     &  763 & M\,87,2,3 & 320 & $(2.1 \pm 0.4)\times 10^{1}$ & 51,74 & 182 & $(4.6 \pm 0.5)$\\
   4406     &  881 & M\,87,2,3 & 143 & $(1.1 \pm 0.2)\times 10^{-2}$ & 50,74 & 115 & $(1.9 \pm 0.5)\times 10^{-3}$\\
   4435     & 1030 & M\,87,2,3 & 133 & $(1.1 \pm 0.2)\times 10^{-2}$ & 50,73 & 147 & $(3.3 \pm 0.4)\times 10^{-3}$\\
   4459     & 1154 & M\,87,2,3 & 106 & $(4.0 \pm 0.9)\times 10^{-3}$ & 92,93 & 49.7 & $(2.2 \pm 0.2)\times 10^{-3}$\\ 
   4472     & 1226 & 4,5,7,8 & 221 & $(3.6 \pm 0.7)$ & 64,83,84 & 17.1 & $(2.5 \pm 0.3)\times 10^{-1}$\\      
   4486     & 1316 & M\,87,2,3,4 & 460 & $(1.2 \pm 0.2)\times 10^{3}$ & 28 & 252 & $(2.0 \pm 0.2)\times 10^{2}$\\
   4526     & 1535 & 4,5,7,8 & 220 & $(3.2 \pm 0.6)\times 10^{-2}$ & 82,103 & 16.1 & $(1.2 \pm 0.1)\times 10^{-2}$\\
   4552     & 1632 & M\,87,1,2,5,6 & 121 & $(1.5 \pm 0.3)\times 10^{-1}$ & 46 & 34.6 & $(6.4 \pm 0.6)\times 10^{-2}$\\
   4621     & 1903 & 1,5,6 & 162 & $(1.5 \pm 0.3)\times 10^{-3}$ & 65,66,85 & 10.9 & $(2.0 \pm 0.2)\times 10^{-4}$\\ 
   4649     & 1978 & 1,5,6 & 216 & $(2.0 \pm 0.4)\times 10^{-1}$ & 65,85,285 & 16.2 & $(3.3 \pm 0.3)\times 10^{-2}$\\
   4660     & 2000 & 5,6 & 207 & $(2.3 \pm 0.1)\times 10^{-3}$ & 280,285 & 11.3 & $(3.4 \pm 0.2)\times 10^{-4}$\\
\hline                                  
\end{tabular}}
\tablefoot{$^*$: NGC\,4365 is not detected at a surface brightness significance of 4$\sigma$ in LOFAR. Column description: (1) NGC entry; (2) VCC entry; (3) LOFAR pointings; (4) LOFAR rms noise in $9''\times{5}''$ map; (5) 144\,MHz flux density; (6) MeerKAT pointings; (7) MeerKAT noise level; (8) MeerKAT flux density.}
\end{table*}

\begin{table*}
\caption{Astrophysical properties of the sample}            
\label{tab:properties}
\setlength{\tabcolsep}{12pt}
\centering                          
\begin{tabular}{l c c c c c c}       
\hline\\[-1.7ex]            
NGC & $\log_{10} M_*$ & $\log_{10} L_{144}$ & LLS & $\alpha^{1284}_{144}$ & $B_{\mathrm{eq}}$\\
 & [$M_\odot$] & [W/Hz] & [kpc] &  & [$\mu$G]\\
\hline\\[-1.7ex]                      
   4262 & 10.20 & $19.37 \pm 0.12$  & <0.68 &  & \\      
   4374 & 11.18 & $23.93 \pm 0.09$ & 13.14 & $-0.69 \pm 0.10$ & 9.20\\
   4406 & 11.22 & $20.55 \pm 0.09$ & <1.37 & $-0.94 \pm 0.15$ & \\
   4435 & 10.30 & $20.52 \pm 0.10$ & <0.73 & $-0.51 \pm 0.12$ & \\
   4459 & 10.78 & $20.04 \pm 0.10$ & <0.98 & $-0.27 \pm 0.11$ & \\ 
   4472 & 11.59 & $23.08 \pm 0.09$ & 149.4 & $-1.21 \pm 0.10$ & 4.13 \\      
   4486 & 11.36 & $25.61 \pm 0.09$ & 65.97 & $-0.82 \pm 0.10$ & 7.80 \\
   4526 & 10.96 & $21.01 \pm 0.12$ & <1.47 & $-0.47 \pm 0.10$ & \\
   4552 & 10.80 & $21.66 \pm 0.09$ & 15.58 & $-0.40 \pm 0.10$ & 3.27 \\
   4621 & 10.28 & $19.42 \pm 0.10$ & <0.65 & $-0.95 \pm 0.15$ & \\ 
   4649 & 11.34 & $21.81 \pm 0.09$ & 33.21 & $-0.82 \pm 0.10$ & 2.82 \\      
   4660 & 10.05 & $19.69 \pm 0.21$ & <0.65 & $-0.86 \pm 0.14$ & \\
\hline                                  
\end{tabular}
\tablefoot{Column description: (1) NGC ; (2) total stellar mass of the host galaxy in stellar units; (3) total $\SI{144}{MHz}$ luminosity (4) largest linear size of the radio emission with upper limits for point and point-like sources, (5) radio spectral index between $\SI{144}{MHz}$ and $\SI{1.28}{GHz}$ defined by $F_{\nu} \propto \nu^{\alpha}$ (6) Equipartition magnetic field strength for the extended sources}
\end{table*}

\subsection{ViCTORIA HBA data}
\label{subsec:lofar}
The ViCTORIA HBA survey \citep{2023A&A...676A..24E} consists of $8\times8$\,h of observations conducted using two parallel beams with a central frequency of $\SI{144}{\mega\hertz}$ spread across nine pointings. The covered area of $\SI{132}{\deg\squared}$ contains the Virgo cluster out to at least $1.75\times$ the virial radius \citep[$r_{\text{vir}}$=3.3$^\circ$,][]{2017MNRAS.469.1476S}.
The data reduction was carried out using a customized and extended version of the strategy that is used for the LOFAR Two-metre Sky Survey \citep{Tasse2021}. After correcting the instrumental systematic effects using observations of bright calibrator sources, a specialised ``{peeling}'' strategy was used. This was necessary to mitigate the dynamic range limitations due to the presence of the extremely bright central source NGC\,4486 (M\,87, also part of our sample). After obtaining a high-quality source model of NGC\,4486 through self-calibration on the source, we accurately calibrated and subtracted the bright source from the visibility data. Subsequently, direction-dependent calibration of the full field could be carried out and the final images were combined to create a mosaic of the field at high ($9''{\times}5''$), low ($20''{\times}20''$) and very low ($1'{\times}1'$) resolution. The median root-mean-square (rms) noise level of the high-resolution mosaic within the virial radius is $140\,\muup{Jy}\,\mathrm{beam^{-1}}$. The systematic uncertainty on the flux density scale was estimated to be 20\%. 

The detailed description of the data reduction and survey properties can be found in \citet{2023A&A...676A..24E}. 
In this previous work, a catalogue of 112 LOFAR-detected Virgo cluster member galaxies was provided based on a $4\sigma$ detection limit in the low-resolution mosaic. For our study, we relax this requirement to sources with a statistical significance of at least $4\sigma$ in either the high- or low-resolution image. This additionally includes NGC\,4262, which is detected at 4.0$\sigma$ in the high-resolution image but only at 3.6$\sigma$ in the $20''$ map, and NGC\,4621, detected at 4.9$\sigma$ in the high resolution image and at 3.8$\sigma$ in the low resolution image.
In total, this leads to 12 galaxies that are part of the ACSVCS and detected at 144\,MHz with LOFAR.
Of the 100 galaxies in the ACSVCS, 6 objects are outside the footprint of the ViCTORIA HBA survey and therefore not part of our analysis. We manually inspected the LOFAR mosaics at the locations of the 82 remaining ACSVCS objects that are covered but not detected by LOFAR to ensure that no sources were missed.

\subsection{ViCTORIA MeerKAT data}
\label{subsec:meerkat}
The ViCTORIA MeerKAT observations are being collected as part of the observing projects MKT22008 in the 2022/23 cycle and MKT23067 in the 2023/24 (PI: de Gasperin) using the L-band receiver system with a central frequency of 1.28\,GHz. A total of 320 pointings cover a 112$\, \mathrm{deg}^2$ region of the cluster in a hexagonal grid with a spacing of $0.58^\circ$. The effective exposure time is 43-45\,min per pointing. For each observation, the telescope cycled between five target sources and the gain calibrator (J1150-0023) nine times in total to maximise $uv$-coverage. In addition, a bandpass calibrator (J1939-6342) and a polarization calibrator (3C286) was included in each observation. The observational campaign and the complete data reduction of this project is ongoing. For this study, we carried out the data reduction of the pointings that cover the 12 sources of our LOFAR-ACSVCS sample and the massive elliptical galaxy NGC\,4365 for which we detect no compact emission in LOFAR.
We consider all pointings that are within $0.4^\circ$ of at least one of the galaxies, this yields 1-3 pointings per galaxy. For NGC\,4365, which is surrounded by low-surface brightness diffuse emission in the LOFAR maps, we include two more pointings to cover the full diffuse emission in LOFAR, this yields four pointings for this galaxy.

For each pointing, the following data reduction steps were carried out: 
Before downloading the data, they were averaged in frequency from an initial channel width of 26.1\,kHz to 0.209\,MHz and channels outside of the frequency range 900--1650\,MHz were removed. The data were then downloaded and further reduced using the 'Containerized Automated Radio Astronomy Calibration' (\texttt{CARACal}\footnote{\url{https://caracal.readthedocs.io/}}) pipeline.
First, the data of the bandpass and gain calibrator were flagged using \texttt{aoflagger} \citep{Offringa2012}, we also flagged channels of known RFI and those  corresponding to the 21\,cm line (1419.8–1421.3\,MHz). Then, the calibrator solutions were derived using a standard \texttt{CASA}-based \citep{McMullin2007} approach and applied to the target field data. These were subsequently averaged to 16\,s in time and 1.67\,MHz in frequency as a compromise between data volume and time- and frequency-smearing effects. Next, we carried out flagging of the target field using the \texttt{tricolour}\footnote{\url{https://github.com/ratt-ru/tricolour}} flagger and then we proceeded with self-calibration.
For the initial model, the field was imaged using \texttt{WSClean} \citep{Offringa2014}. Then, four rounds of self-calibration were carried out using the \texttt{cubical}\footnote{\url{https://caracal.readthedocs.io}} solver and \texttt{WSClean}. In the first round, we solved only for diagonal phase-solutions on 128\,s timescales and 3.34\,MHz frequency windows. For the further iterations, we solved also for diagonal amplitude solutions (one solution per scan and per 3.34\,MHz window), while increasing the frequency resolution of the phase solutions to one solution per channel. 

For pointings in proximity of the extremely bright source NGC\,4486 (= M\,87\,/\,Virgo\,A), we assured that the image region contained the bright source and applied a manual mask containing NGC\,4486 during deconvolution. For pointing 46, we needed to additionally peel NGC\,4486 from the $uv$-data. For this, we phase-shifted the observation to NGC\,4486, carried out ten cycles of self-calibration of the source using a manual cleaning-mask. Then, the final model found for M87 was corrupted with the final calibration solutions and subtracted from the data set. Subsequently, the data were phase-shifted back to the original phase-center and self-calibrated identical to the other observations.

After self-calibration, the final images were corrected for the effect of the primary beam and neighbouring pointings were combined to a mosaic. To achieve this, a primary beam-corrected image was created for each self-calibrated image, the area below a primary beam attenuation factor of 30\% was disregarded. Afterwards we combined the neighbouring pointings by convolving the primary-beam corrected images to the smallest common circular synthesized beam (between $9.7''$ and $12.7''$) and re-gridded the images to a common pixel layout. We obtained the final mosaics by taking the average between the overlapping pixels of the individual pointings. The weights were calculated according to the central noise level in the pointing image and the primary beam attenuation factor at the location of each pixel. In total, we created seven mosaics which cover one to three sources of the sample. 

For the flux density scale of the MeerKAT maps, we conservatively assume a systematic uncertainty of 10\% \citep[e.g.][]{Knowles2022}.
We verified that this assumption is reasonably by comparing the flux density of compact sources between the MeerKAT mosaic to the NRAO VLA Sky Survey (NVSS, \citealt{1998AJ....115.1693C}).

\subsection{Source subtraction}
\label{subsec:source-sub}
Two of the ACSVCS galaxies (NGC\,4365 and NGC\,4472) show particularly extended diffuse emission in the LOFAR HBA images which is not observable in the MeerKAT images at the nominal resolution of ${\approx}10''$. To check for the presence of low-surface brightness diffuse emission in MeerKAT data, we performed a source subtraction and low-resolution imaging procedure for these targets. 
For this, each pointing that covers the two sources as listed in Table \ref{tab:image_prop} was imaged using \texttt{WSClean}. Here, a $uv$-cut of 3000$\lambda$, where $\lambda$ is the wavelength, was employed to obtain an image containing only compact sources (${\sim}1'$ or smaller). From this image, we created a compact-source mask. The mask regions corresponding to the central AGN in NGC\,4365 and NGC\,4472 were manually removed so that also possible components corresponding to the core of these galaxies will be included in the low-resolution image. A repeated run of \texttt{WSClean} with the same $uv$-cut but using the compact-source mask was used to create a model of the compact sources in the image (excluding NGC\,4365 and NGC\,4472). The visibilities corresponding to this model were subtracted from the calibrated $uv$-data of each pointing. This subtracted data was imaged, this time tapering the imaging weights such that the resulting image resolution is $1'$. This way, we recover a low-resolution image of only the extended emission. Lastly, we corrected the low-resolution images for the primary beam and mosaicked them following the procedure described in Section \ref{subsec:meerkat}.

\subsection{XMM-Newton data}
\label{subsec:xmm-newton}
In this work, we reanalysed three XMM-Newton observations of M49 (ObsID: 0761630101, 0761630201, 0761630301), originally presented in \cite{2019AJ....158....6S}. All observations were taken in January 2016, with a total effective exposure time of $\sim240$\,ksec. Data analysis were performed using XMM-Newton Science Analysis System (SAS) version (xmmsas\_20210317\_1624-19.1.0). For details of the data reduction procedure, we refer to \cite{2019AJ....158....6S}. Given that MOS1 CCDs 3 and 6 were lost to micrometeorites, only MOS2 and pn were used for imaging analysis but all the data were included in spectrum analysis. We first obtain an exposure and vignetting corrected EPIC image in the energy band of 0.7--1.3\,keV with the instrumental background subtracted. After removing point sources, we derived its surface brightness profile and fit it with a double-$\beta$ profile model. We obtain a residual image after dividing the XMM-Newton image by an azimuthally symmetric image generated from the best-fit double-$\beta$ model.         
To obtain a deprojected pressure profile of the ICM, we utilize the south region in between the two radio tails that we consider to be less affected by stripping or AGN bubbles. We extracted spectra from 9 sectional annuli 
from the centre of M49 to 96\,kpc. We fit each region to an absorbed thermal emission model, {\tt phabs}$\times${\tt apec}, in XSPEC (version 12.11.1). 
The temperature, metallicity, and normalisation were allowed to vary freely. 
The parameters of the astrophysical background (local bubble, Milky Way, and cosmic X-ray background) were taken from \cite{2019AJ....158....6S}, determined with offset XMM-Newton pointings. Instrumental backgrounds were modelled with a set of fluorescent instrumental lines \citep[as listed in Table~2 in][]{2017ApJ...851...69S} plus a broken power law with a break at 3\,keV. 
We derive its 3D density profile, $\rho(r)$, assuming its shape follows the deprojection of its surface brightness profile. We then adjust its amplitude such that the corresponding normalisation profile best match the one obtained in the spectrum analysis (see \cite{2019AJ....158....6S} for detail).
After obtaining $\rho(r)$, we fit the projected temperature profile to 
\begin{equation}
T_{\rm 2D}=\frac{\int \rho^2T_{\rm3 D}^{1/4}dz}{\int \rho^2T_{\rm3 D}^{-3/4}dz},
\end{equation}
where z is along the line of sight and for which we assume the deprojected temperature profile, $T_{\rm 3D}$, follows a functional form applicable to cool core clusters \citep[Eq.1 in][]{2019AJ....158....6S}. 
We obtain the 3D pressure profile for the quiescent south region of the ICM by multiplying the 3D temperature and density profiles.

\section{Radio analysis of the sample}
\label{sec:full sample}
While the size of our radio-detected sample is limited, it contains the volume of the Virgo cluster, the largest aggregation of early-type galaxies in the nearby Universe. Therefore, we are able to sample AGN with a wide range of radio luminosities  (more than six orders of magnitude, from NGC\,4262 with $\SI{2.3e19}{\W\per\Hz}$ to  NGC\,4486 with $\SI{4.1e25}{\W\per\Hz}$). In this section, we investigate the radio properties of the objects in our sample and explore their connection with the host galaxy parameters.

\subsection{Radio morphology}
Five out of the twelve galaxies in our sample have $3\sigma_\mathrm{rms}$ radio contours that are much larger than the beam size of the image. These are NGC\,4374, NGC\,4472, NGC\,4486, NGC\,4552 and NGC\,4649. We define these objects as extended sources and will refer to them as such in the following. They form the most radio luminous galaxies in our sample with $\log_{10} L_{144} \geq 21.5$. All of the extended sources show clear indications of the presence of jets. For NGC\,4472, our LOFAR maps reveal for the first time large and diffuse tails extending over $\SI{150}{kpc}$, suggesting restarted activity of the central AGN. This source is analysed in detail in Section \ref{subsec:ngc4472}.

The other seven sources in the sample show 144\,MHz radio emission close to the resolution limit of our LOFAR maps ($9''{\times}5''$). These include three sources showing marginal extension (NGC\,4406, NGC\,4459 and NGC\,4526) which could be either due to the intrinsic size of these sources or alternatively, due to smearing effects in the images related to residual phase errors and the finite time- and frequency resolution of our data.

To analyse whether these sources are truly marginally resolved, we created a source model of a 0.25 square degree sky area around the objects using \texttt{PyBDSF} \cite{2015ascl.soft02007M}. The reconstructed full width at half maximum (FWHM) extension of all three sources is at least 8$''$ larger than the resolution limit in both the major and minor axis. We therefore compared them to the extension of other point sources in the vicinity by considering the ratio of the peak and total flux. For truly unresolved sources in an image with negligible systematic effects, this ratio should approach 1 \citep[e.g. Figure 6 in][]{2022A&A...659A...1S}. The presence of residual systematic effects in the data may introduce an artificial smearing of point sources which should affect point sources of the same flux density in a similar manner. 
The position of the three sources, however, does not conclusively show whether they are truly point sources or not. In comparison by eye the MeerKAT images of NGC\,4406 and NGC\,4526 show a similar ambiguity in their extent as seen in the LOFAR images. This is not the case for NGC\,4459 which is only slightly more extended than the resolution of the image. However, if the extent of the radio emission would be of intrinsic nature, there is a chance that it could be old emission not visible anymore in the MeerKAT observations, making it appear as a point source there.

As we cannot resolve this ambiguity and because we wish to have a clear distinction from the other four sources clearly below the resolution limit, they will be referred to as point-like sources and point sources respectively.

The LOFAR HBA maps of all sources and the MeerKAT maps of the eleven covered sources are shown in Appendix \ref{sec:app_lofar} and Appendix \ref{sec:app_meerkat} respectively with their half light radius and distance to the central galaxy NGC\,4486 marked.

\subsection{Radio emission and  stellar mass of host}

\begin{figure}[h]
     \centering
     \includegraphics[width=0.95\columnwidth]{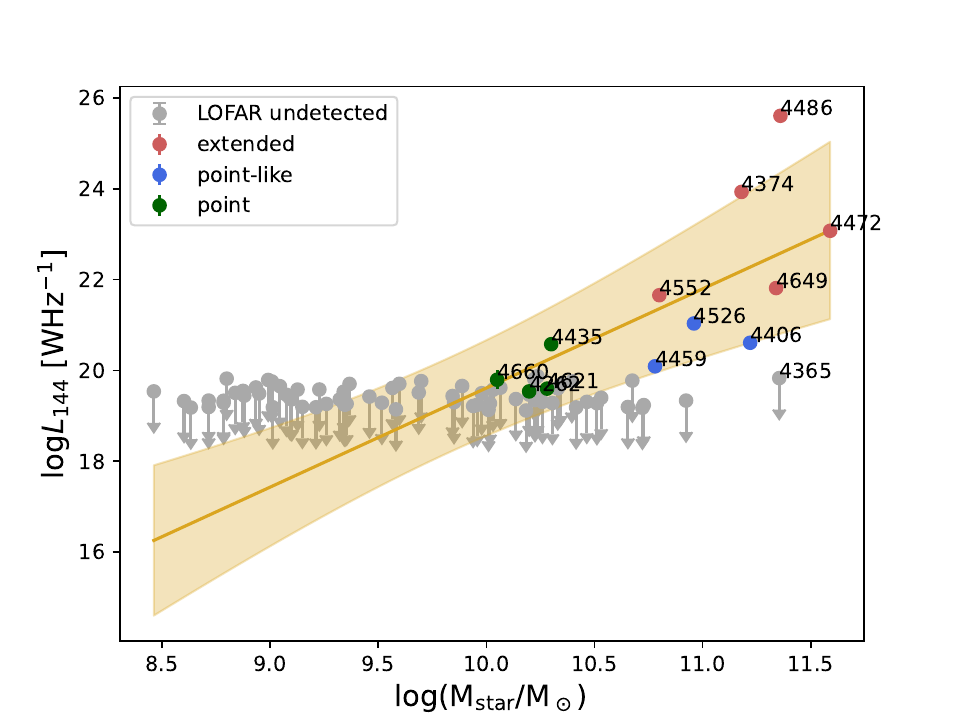}
     \caption{Stellar mass against the total radio luminosity for all galaxies in the sample. The different colours in the plot mark their morphological classification: Extended sources are shown in red, point-like sources in blue and point sources are coloured green. The grey points are upper limits derived for the galaxies which are not detected at 144\,MHz 4\, $\sigma$. The yellow line shows the best-fitting relation and its 1$\sigma$ uncertainty band.}
     \label{fig:mass_lum}
\end{figure}

In Figure \ref{fig:mass_lum} we show the $\SI{144}{MHz}$ luminosity of the sources in the sample plotted against the stellar mass of their host galaxies. For the point and point-like sources, the radio flux densities were measured from the high-resolution maps. For the extended sources, we instead used the low-resolution maps to ensure that the diffuse emission is properly deconvolved. The flux density upper limits of the 144\,MHz-undetected sources were included by taking the 4\,$\sigma$ noise of the residual images at low resolution close to the position of the sources. For the uncertainty of the flux measurement, both the rms noise $\sigma_{\mathrm{rms}}$ and the uncertainty of the flux density scale of 20\% for the LOFAR HBA maps were accounted for. The flux uncertainty is given by:
\begin{equation}
    \Delta S = \sqrt{\sigma_{\mathrm{rms}}^2 + (0.2 \times S)^2}
\end{equation}
where $S$ is the measured flux.
We calculated the luminosity using $L = 4\pi d_L^2S$ where we used the distances determined with the surface brightness fluctuation method in \cite{2009ApJ...694..556B}. Due to the close proximity of the Virgo cluster we neglect $K$-correction. The uncertainties of the luminosities were then calculated from the flux uncertainties. The stellar masses of the host galaxies are taken from \cite{2008ApJ...681..197P}.

Taking the LOFAR upper limits into account, the sample shows a positive correlation as given by the Kendall's $\tau$ correlation coefficient: $\tau = 0.37$ with a p-value of $\mathrm{p} = 1.2 \times 10^{-6}$ (if only the detected sources are considered Kendall's tau becomes larger with  $\tau = 0.70$). 
We perform a linear fit to the full sample (including the upper limits) in log space using the Bayesian LInear Regression in Astronomy method \citep[LIRA;][]{2016MNRAS.455.2149S} of the following form:
\begin{equation}
\log_{10} \left( \frac{L_{144}}{10^{19.7}\,\mathrm{W\,Hz}^{-1}} \right) =  A \log_{10} \left( \frac{M_{\mathrm{star}}}{10^{9.8}\,\mathrm{M}_{\odot}} \right) + B,
\end{equation}
where $A$ and $B$ are the slope and intercept of the correlation, respectively.
We find that the radio luminosity of the AGN shows a steep correlation with the mass of their host galaxies. Our best fit parameters for the slope and intercept are:
\begin{equation}
    \log_{10} \left( \frac{L_{144}}{10^{19.7}\,\mathrm{W\,Hz}^{-1}} \right) = (2.2 \pm 0.9) \log_{10} \left( \frac{M_{\mathrm{star}}}{10^{9.8}\,\mathrm{M}_{\odot}} \right) - (0.5 \pm 1.0).
\end{equation}

A connection between the radio luminosity and the stellar mass of a galaxy is not unexpected, due to the scaling of the stellar mass with the black hole mass, which in turn scales with its radio emission \citep{2006MNRAS.365...11C}. This connection has been found and studied in other works as for example in the work of \cite{2010A&A...511A...1B}. There they found that a correlation between the radio luminosity and host galaxy mass only holds up to $\log_{10} L_{\mathrm{1.4}} = \SI{23.5}{\watt\per\hertz}$. To check whether our observations agree with those of \cite{2010A&A...511A...1B}, we rescale their threshold (calculated for luminosities at $\SI{1.4}{\giga\hertz}$) to a frequency of $\SI{144}{\mega\hertz}$, assuming a mean spectral index of $\alpha = -0.7$. The threshold corresponds to $\log_{10} L_{\mathrm{144}} = 24.2$. Only one source in our sample is brighter than the threshold luminosity. Therefore the found correlation is in agreement with the findings of \cite{2010A&A...511A...1B}.

Another variation is the correlation between the fraction of galaxies above a certain radio luminosity and the mass of the host galaxy. This has been studied in many other publications \citep[e.g.][]{2005MNRAS.362...25B,2007MNRAS.375..931M,2009AJ....138.1990C,2019A&A...622A..17S}, where a strong correlation between the two parameters is found. Such a correlation should be visible as well in a plot of the radio luminosity against the mass and can be seen in our data.

\cite{2019A&A...622A..17S} was able to precisely measure the relation between the fraction of radio-loud galaxies and the host mass and extended it to lower luminosities. They additionally find that all galaxies with $M_* > 10^{11}M_{\odot}$ are always showing nuclear activity at a level of $L_{150}>10^{21}\,\mathrm{W\,Hz^{-1}}$. This result is in agreement with later studies by \cite{2022ApJS..258...30G} and \cite{2022A&A...660A..93C}, where they find that all sources above the mentioned mass are at least detected by LOFAR. In the case of \cite{2022A&A...660A..93C} not all of the galaxies host an active AGN. Some are remnant sources still visible at lower frequencies.

Interestingly, NGC\,4365, one of the galaxies in our sample, does not fit this scenario. Despite its high mass of $M_* > 10^{11}M_{\odot}$, it does not meet our requirement of emitting at a 144\,MHz surface-brightness significance of 4$\sigma$ in either the high- or low-resolution LOFAR maps (corresponding to a $4\sigma$ limit of $L_{144}<6.75\times 10^{19}\,\mathrm{W\,Hz^{-1}}$). This indicates that the galaxy is currently in a phase of particularly low activity. A more detailed analysis of the source itself and a possible detection of a previous significantly stronger phase of activity are presented in Section \ref{subsec:N4365}. 

\subsection{Radio luminosity and largest linear size}
\begin{figure}[h]
     \centering
     \includegraphics[width=0.95\columnwidth]{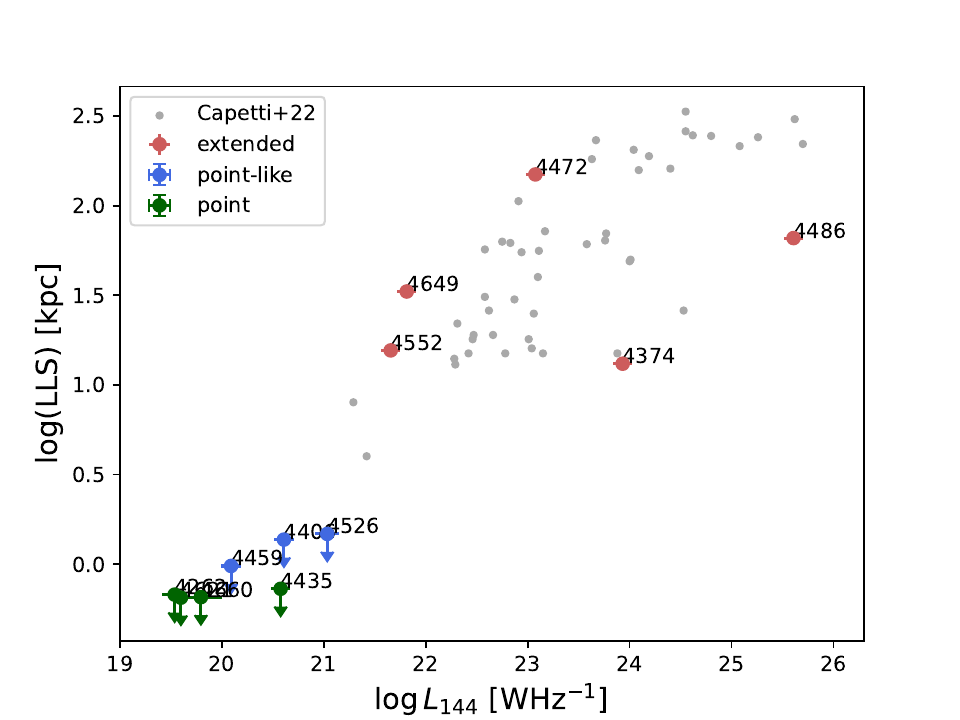}
     \caption{Largest linear size against the total radio luminosity for all radio detected galaxies in the sample. The sources are colour-coded the same way as in Figure \ref{fig:mass_lum}. Only upper limits in LLS are given for the point and point-like sources. The smaller light grey points are data from \cite{2022A&A...660A..93C} for a sample of nearby AGN.}
     \label{fig:las_lum}
\end{figure}

Figure \ref{fig:las_lum} shows the largest linear size (LLS) of the sources in dependence of their $\SI{144}{MHz}$ radio luminosity. Like the radio flux of the sources the LLS has been measured in different images for the extended sources and point sources. For the extended sources we used the low resolution radio images to determine the extent of the emission. For the point sources it is only possible to determine an upper limit for the extent of the sources in the high resolution images. We set the linear size for the four point sources as the resolution limit of the image scaled to the distance of the source. As the radio emission of the three point-like sources seems to extend beyond the resolution limit, we conservatively set the upper limit to the FWHM major axis that we obtained with the \texttt{PyBDSF} source finder (which are 31.6'' for NGC\,4406, 25.2'' for NGC\,4459 and 36'' for NGC\,4526).

We can see a positive trend between the linear size of the radio emission (including the upper limits of the compact sources) and the radio luminosity which is reaffirmed by the correlation coefficient and p-value ($\tau = 0.561$, p-value = $0.009$) for our sample. Such a behaviour is consistent with the one found in other samples like \cite{1990A&A...227..351D} or \cite{2022A&A...660A..93C}. Especially the sample of \cite{2022A&A...660A..93C} has similar conditions as our sample, consisting of radio sources in nearby early-type galaxies and being observed with LOFAR. In Figure \ref{fig:las_lum} we can see that the two samples match in their distribution, confirming a connection between the extent of the radio emission and the luminosity of the sources. By including point sources in this relation which were not considered in \cite{2022A&A...660A..93C}, we are able to extend the analysis to lower luminosities.
The point sources show lower luminosities than the extended sources, taking the sample of \cite{2022A&A...660A..93C} into account.  

\subsection{Linear size and spectral index}
\begin{figure}[h]
     \centering
     \includegraphics[width=0.95\columnwidth]{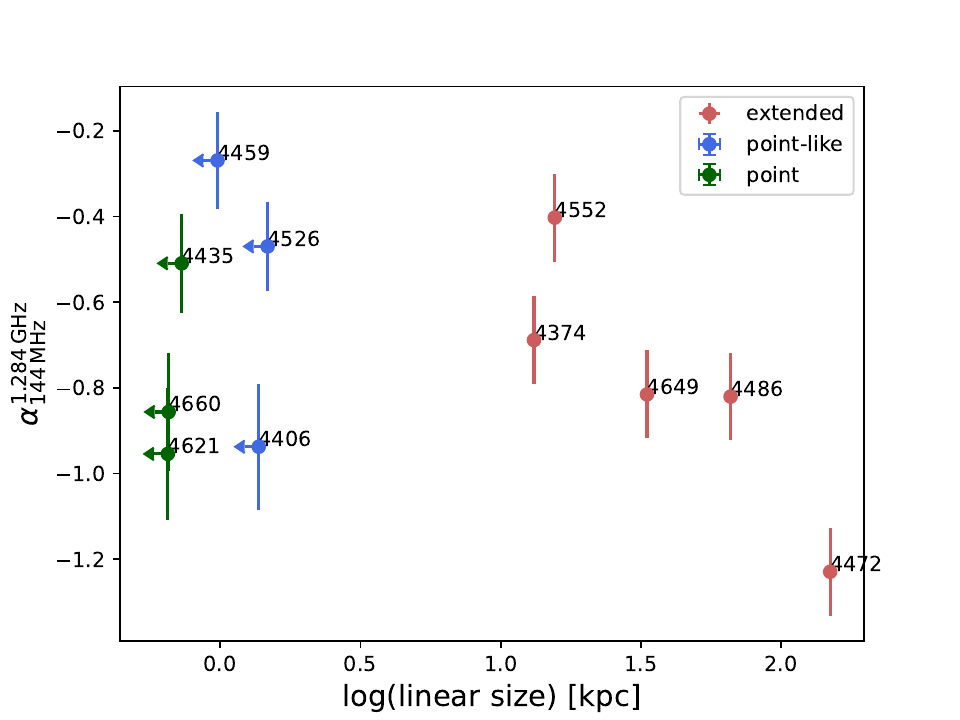}
     \caption{Largest linear size against the total spectral index for 11 of the 12 galaxies which were covered by MeerKAT. The spectral index was calculated using $\SI{1.28}{GHz}$ and $\SI{144}{MHz}$ flux measurements. Only upper limits in LLS are given for core-like radio emission. We can potentially see a trend for the extended sources.}
     \label{fig:las_spec}
\end{figure}

In Figure \ref{fig:las_spec} we compare the linear size of the radio sources with their spectral index between $\SI{144}{MHz}$ and $\SI{1.28}{GHz}$\footnote{The spectral indices $\alpha$ are defined as $F_{\nu} \propto \nu^{\alpha}$.}. To determine the spectral index the LOFAR and MeerKAT images were convolved and regridded to match in their beam size. Then the radio flux was measured in these images using the same region as defined by the emission seen in the LOFAR images for each source. With this we determined a spectral index for eleven out of the twelve sources, as one source was not covered by the MeerKAT observations. Due to the availability of observations in two different frequency bands, we get a solid estimation of the spectral index.

In the plot we can see a general trend of the extended sources having a flatter spectral index with smaller linear size. This does not seem to hold true for the point and point-like sources. Here, we can see a spread in the spectral indices from $\alpha_{144}^{1284} \approx -1$ to  $\alpha_{144}^{1284} \approx -0.2$. A similar spread in the spectral index of  sources which are unresolved on scales between 1.2 and 15 kpc has been found in the work of \cite{2022A&A...660A..93C}. The spectral slopes in their sample show a spread of $-1.2 < \alpha^{1400}_{150} < 0.4$, larger than the one we find. In their sample, however, a larger number of point sources across different environments and distances have been considered, which are likely to produce a larger spread in spectral slopes.

In general we would suspect some kind of connection between the spectral index of radio sources and the extent of the radio emission, as AGN with older radio emission, as implied by the steeper spectral index, have more time for the jets to expand into the surrounding medium than sources with relatively new radio emission. Such a trend has been found for extended radio galaxies by \cite{2024arXiv240308037S}. Point sources are additionally strongly impacted by synchrotron self-absorption which flattens the spectral index. So we would expect their spectral index to be generally flatter than the spectral index of extended sources. 

The visible radio emission of the AGN jets, however, is also affected by the environment, the evolution and interaction of the host galaxies and the orientation of the jets towards the observer, all of which can affect the spectral index and extent of each source and therefore cause a scatter in any possible correlation between them. Additionally it is implied that there is a generally different behaviour of the jets in point sources and extended sources. \cite{2022A&A...660A..93C} find that a majority of the radio sources in early-type galaxies are smaller than 4$\, \mathrm{kpc}$. They argue that this implies that the size distribution of radio emission is not determined by age alone but by different jet properties of extended sources and point sources, with the point sources possibly produced by slower jets.

We speculate that a relation between the size of a radio source and the spectral index may only be visible for extended sources. A similarity of the environment of the sources may reduce the scatter in such a potential correlation. In our case all galaxies are part of the Virgo cluster and the four extended galaxies, with the exception of NGC\,4486, show local gas densities of the ICM in the range of $\SI{1e-5}{\per\cm\cubed}$ to $\SI{1e-4}{\per\cm\cubed}$ and are therefore very similar. The local gas densities have been estimated from the gas density model of the Virgo cluster by \cite{2024arXiv240117296M}.

All the sources we consider extended are resolved and have a radio emission above the resolution limit, making them detectable in their size and spectral index.
For point sources we only can determine an upper limit for the size and the possibility exists that there is extended radio emission present which is well below the detection limit. This makes them a more uncertain sample that could also due to the possibility of different jet behaviour, have a different or no connection between the spectral index and size.

We additionally analyse six out of the seven compact sources for which we have MeerKAT observations  available regarding their spectral shape. To do this, we additionally use the \SI{8.4}{\giga\hertz} VLA observations from \cite{2009AJ....138.1990C} available for five sources in our sample (NGC\,4406, NGC\,4435, NGC\,4459, NGC\,4526 and NGC\,4660) and the \SI{8.5}{\giga\hertz} VLA observation for NGC\,4621 from \cite{2008ApJ...675.1041W}. With the radio fluxes at three different frequencies available we are able to calculate the lower spectral index using 144\,MHz and 1.28\,GHz data and a higher spectral index using 1.28\,GHz and 8.4\,GHz data. Comparing those two spectral indices gives us information about the radio spectral shape.
\begin{figure}[h]
	\centering
	\includegraphics[width=0.9\columnwidth]{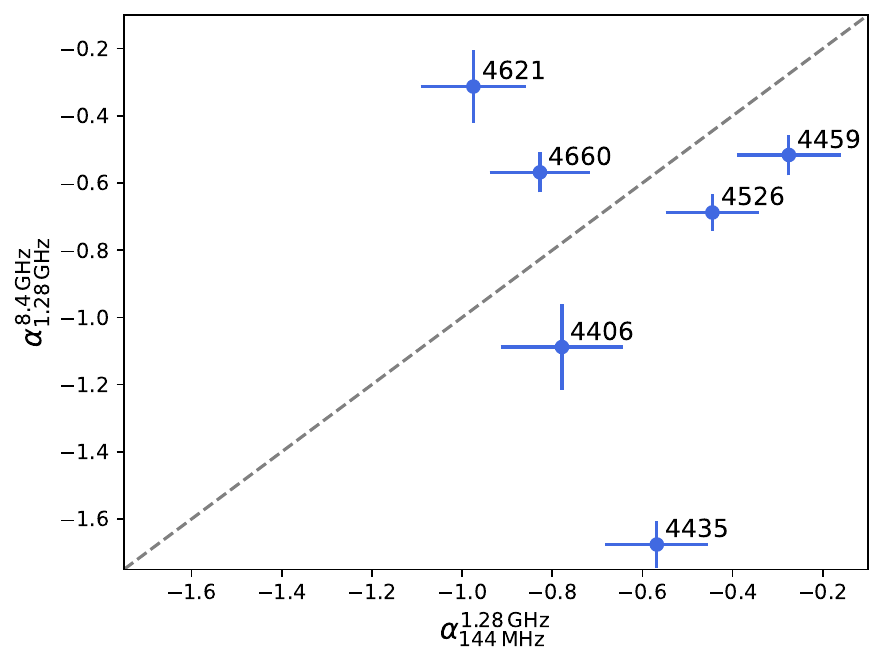}
	\caption{Colour-colour plot for the compact sources. The grey dotted line corresponds to a power-law spectra.}
	\label{fig:point_spec}
\end{figure}

Figure \ref{fig:point_spec} shows the two spectral indices of each source plotted against each other. If those would be the same the sources would be situated on the grey dotted line and follow a power-law spectrum. A position below this line indicates a flatter spectral index at lower frequencies, which is the case in self-absorbed and in spectrally aged sources. For our sample four sources are situated in that regime. NGC\,4459 and NGC\,4526 show signs of synchrotron self-absorption due to their flat lower spectral index. NGC\,4406 has a steep spectral index in both high and low frequency which is in agreement with a spectrally aged source. NGC\,4435 shows the strongest deviation of the power-law slope with a very steep $\alpha^{8.4}_{1.28}$ in comparison to the flatter $\alpha^{1.28}_{0.144}$. This could be an indication of a strongly aged source. However, for this source \cite{2009AJ....138.1990C} found hints of elongation at 8.4\,GHz which are not visible in our LOFAR and MeerKAT observations at lower resolution. This could lead to a lower core flux measured for this source at 8.4\,GHz, producing a steeper spectral index.
Two compact sources in our sample are situated above the injection power-law line, indicating a flattening of the spectral index towards higher frequencies. One of those sources is NGC\,4621. For this source, \cite{2008ApJ...675.1041W} found an indication for variability in flux measurements at the same frequency but six years apart. The calculation of a spectral index using observations made at different time may therefore not produce a reliable spectral index and the position of the source in the colour-colour plot may be a consequence of this variability.

\section{Properties of the extended sources}
\label{sec:extended sources}
In this part of our analysis we take a closer look at the five extended sources in our sample. We will especially analyse the radio emission of the galaxy NGC\,4472 and its effect on the surrounding ICM. We also regard the remnant source NGC\,4365 for which no compact emission above the detection limit in LOFAR is observed.

\subsection{Spectral index maps}
\label{subsec:specidx}
\begin{figure*}[h]
    \centering
    \begin{subfigure}[h]{0.6\columnwidth}
        \centering
        \includegraphics[width=\linewidth]{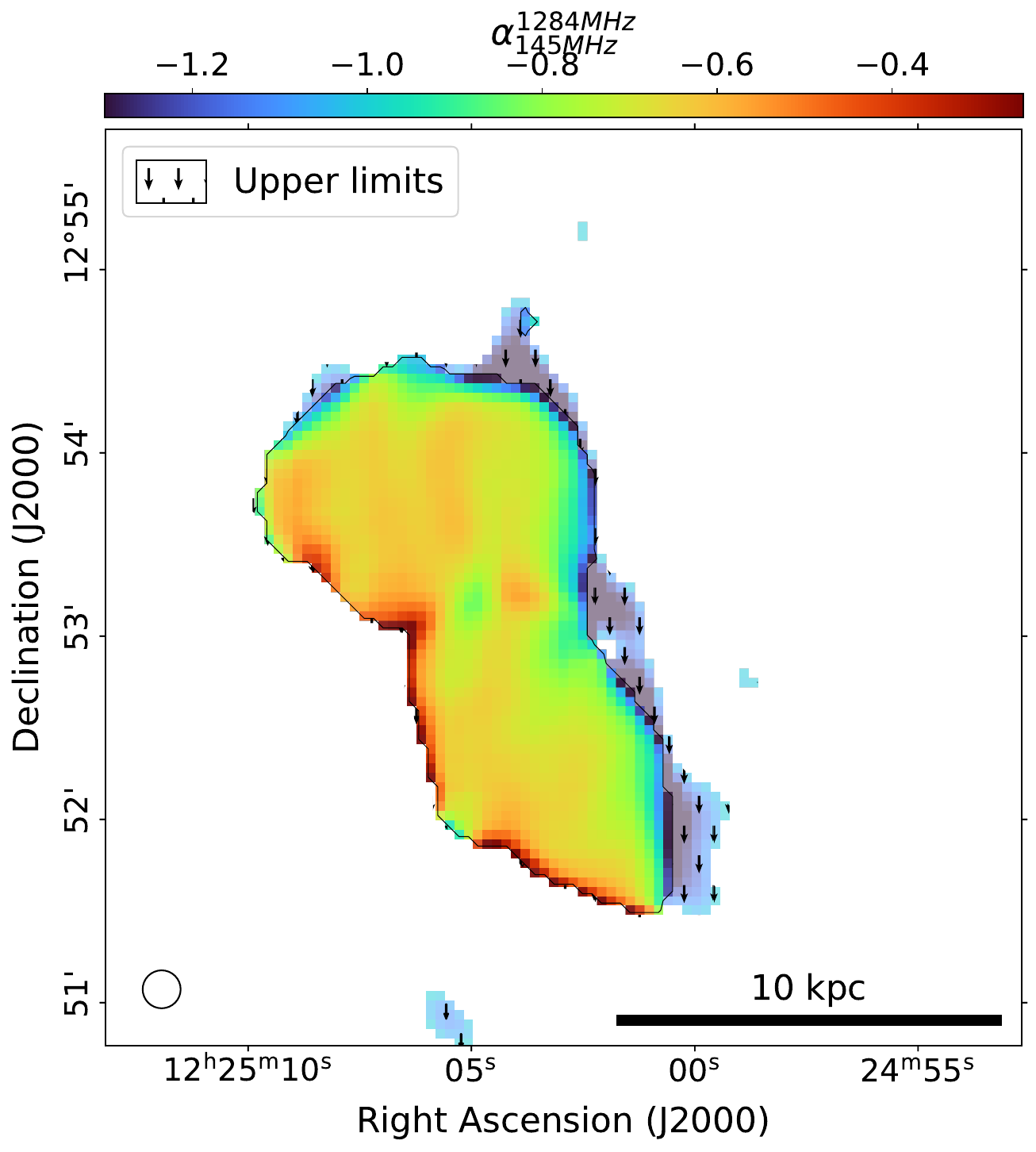}
        \caption{NGC\,4374}
        \label{fig:spidx_4374}
    \end{subfigure}
    \begin{subfigure}[h]{0.62\columnwidth}
        \centering
        \includegraphics[width=\linewidth]{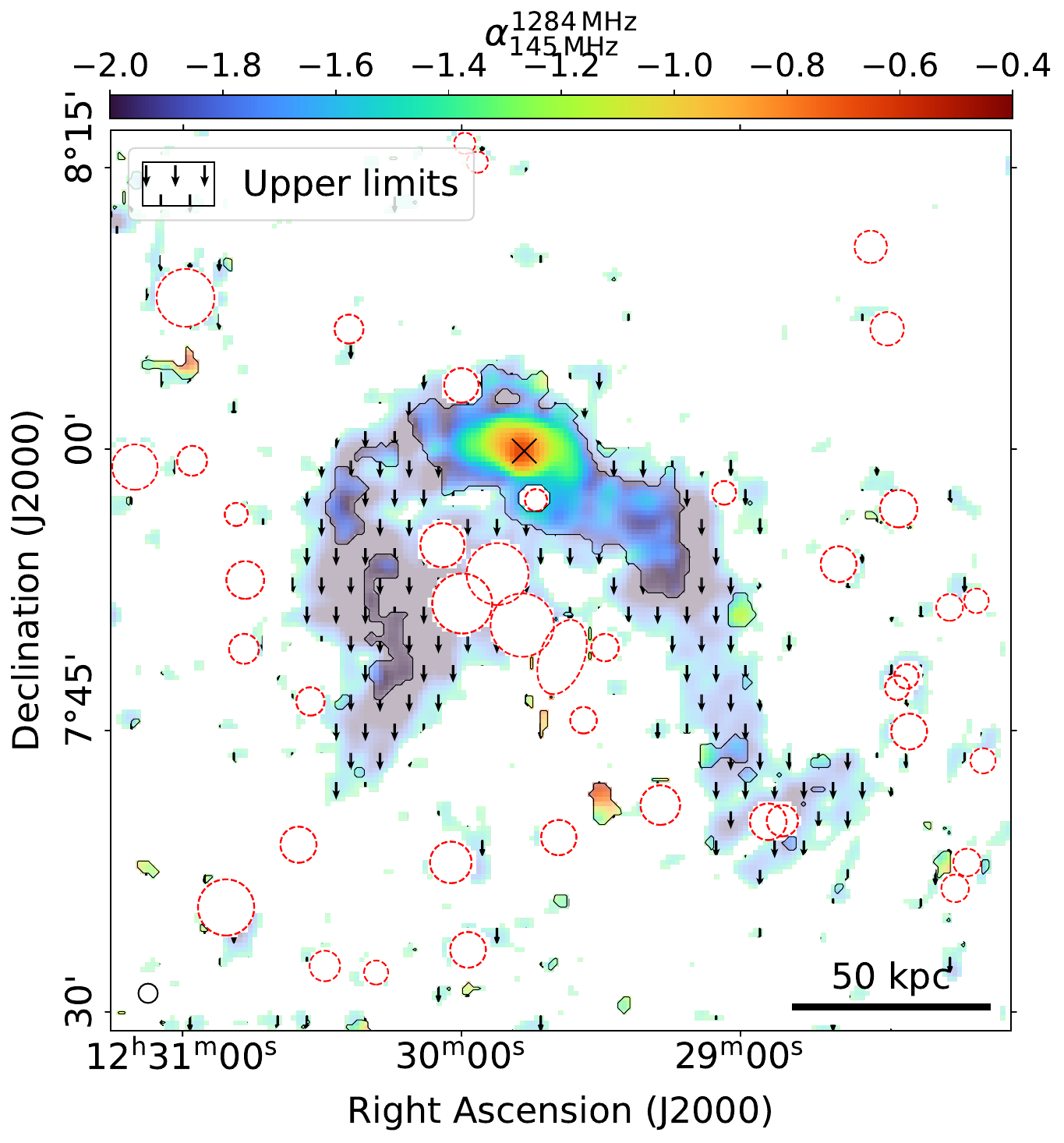}
        \caption{NGC\,4472}
        \label{fig:spidx_4472}
    \end{subfigure}
    \begin{subfigure}[h]{0.63\columnwidth}
        \centering
        \includegraphics[width=\linewidth]{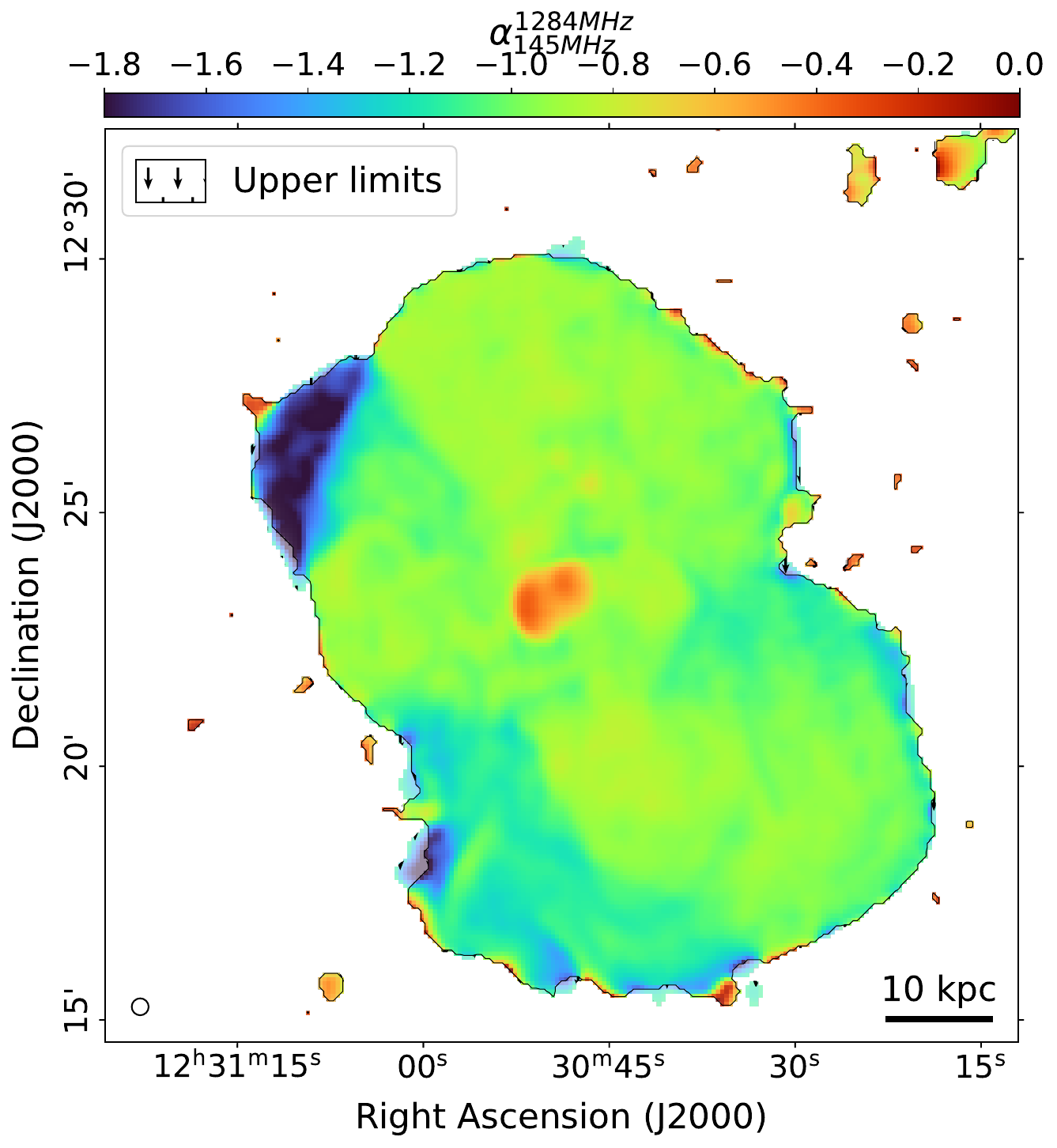}
        \caption{NGC\,4486}
        \label{fig:spidx_4486}
    \end{subfigure}
    \begin{subfigure}[h]{0.63\columnwidth}
            \centering
            \includegraphics[width=\linewidth]{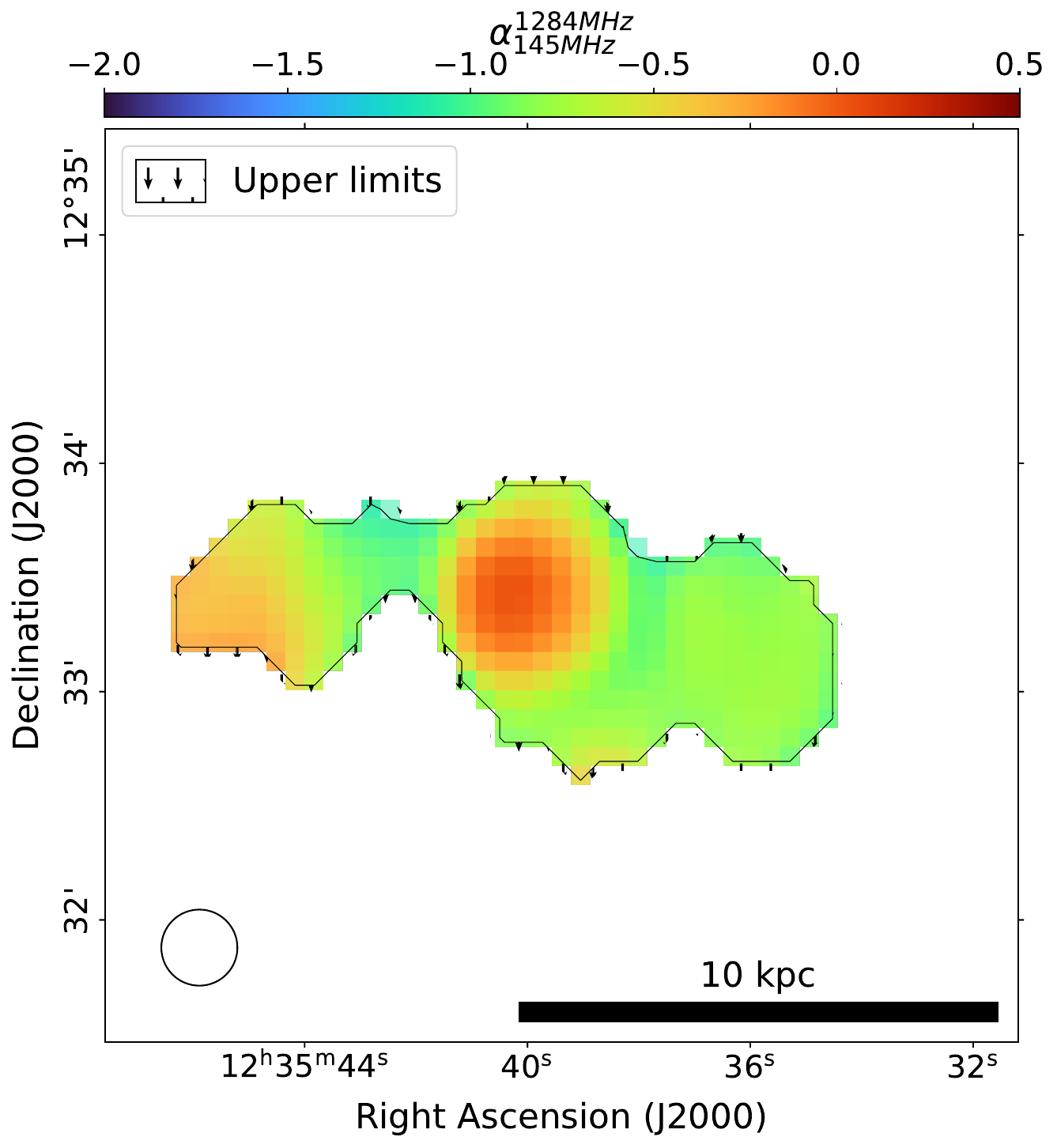}
        \caption{NGC\,4552}
            \label{fig:spidx_4552}
    \end{subfigure}
    \begin{subfigure}[h]{0.63\columnwidth}
            \centering
            \includegraphics[width=\linewidth]{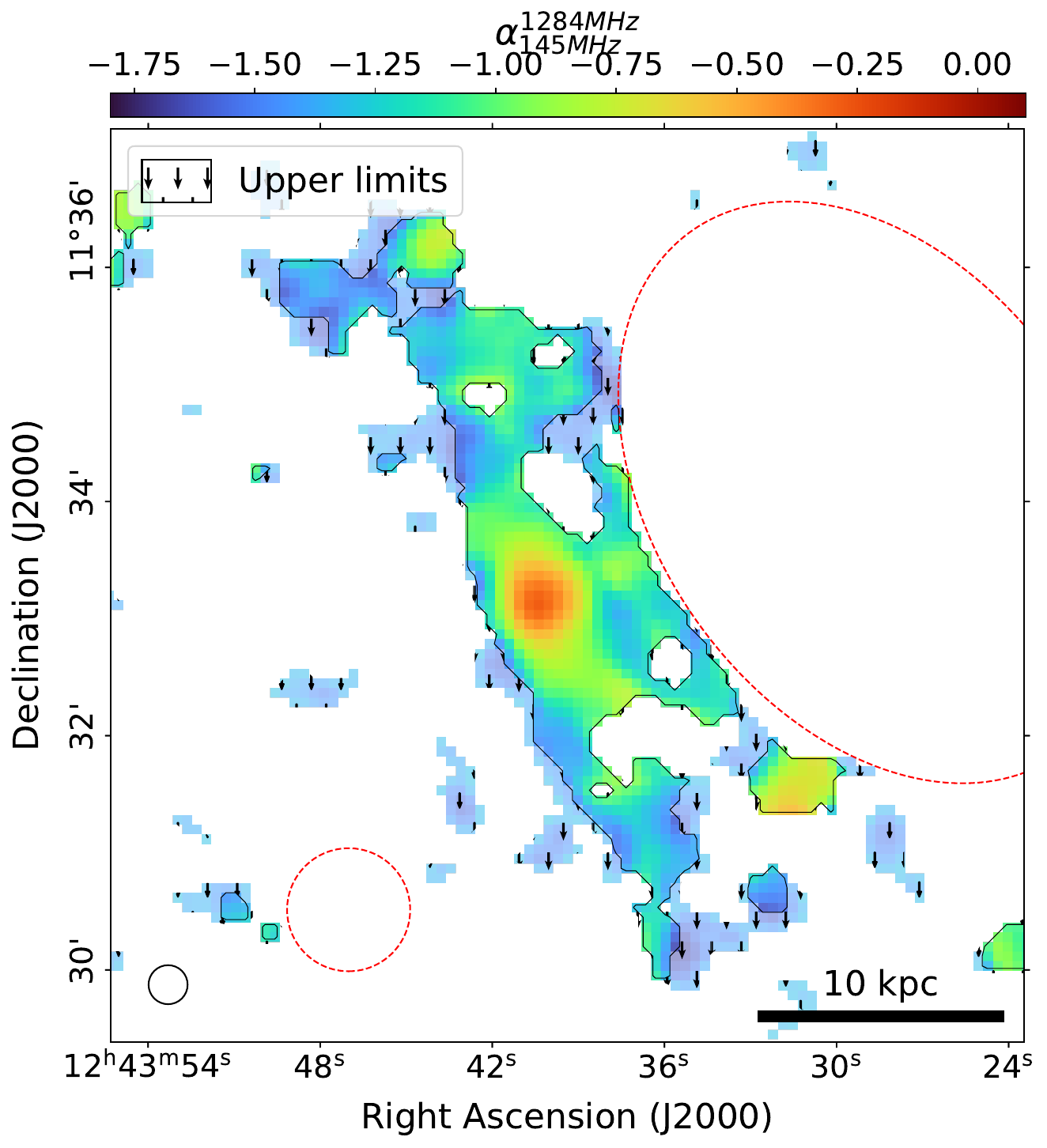}
        \caption{NGC\,4649}
            \label{fig:spidx_4649}
        \end{subfigure}
    \caption{Spectral index maps of the five galaxies with extended radio emission. Dashed red circles mark unrelated sources that were masked. The maps are created using pixels above 3$\sigma$ significance in both the \SI{144}{MHz} LOFAR and \SI{1.28}{GHz} MeerKAT measurements at the same circular synthesized beam size of $20''$, as indicated by the circle in the lower left corner. The colour scales are different in each image. Upper limits of the spectral index are indicated by downward pointing arrows. For the low surface-brightness emission of NGC\,4472, we instead use the LOFAR compact source-masked and the MeerKAT compact source-subtracted images at $1'$ resolution and display pixels above a statistical significance of $2.5\sigma$ in both images.}
    \label{fig:spidx_maps}
\end{figure*}

To create spectral index maps for the five extended sources in the sample, we convolved the low-resolution LOFAR and MeerKAT images to match in their synthesised beam size. This results in a resolution of $20''{\times}20''$. For the extended source NGC\,4472 the spectral index map has a resolution of $62''{\times}62''$ due to the use of the source subtracted MeerKAT image, produced with the process explained in Section \ref{subsec:source-sub}. The spectral index is calculated for each pixel where the emission is above a detection threshold of 3$\sigma$ in both images. For NGC\,4472 the detection threshold has been lowered to 2.5$\sigma$ to highlight the low-surface brightness diffuse emission of this galaxy. Pixels for which a spectral index could be calculated are surrounded by a black contour. In parts where only the LOFAR emission is above the detection threshold, the corresponding upper limit for the spectral index was calculated. These regions are marked with downwards-pointing arrows. The spectral index maps are displayed in Figure \ref{fig:spidx_maps} and the corresponding uncertainty maps can be found in \autoref{sec:app_spidxerr}.

The lack of visible radio emission in MeerKAT images, where only an upper limit for the spectral index has been calculated points towards the presence of radiatively old radio emission which experienced spectral steepening. Most prominently this can be observed for NGC\,4472. Here, a wide-angle tail extending over multiple hundreds of kiloparsecs is at this resolution only present in the LOFAR observations of our data. This results in steep spectral indices in these tails of $\alpha \approx -1.6$. A more detailed analysis of the radio tails of NGC\,4472 is given in Section \ref{subsec:ngc4472}.

For the five galaxies with extended radio emission we can see a flatter spectrum towards the centre with $\alpha > -0.6$. We would expect such a flattening in the core due to the presence of synchrotron self-absorption. This effect can be strong enough to create an inverted spectral index, as seen for NGC\,4552.

\subsection{Equipartition magnetic fields}
For the five extended sources we calculated the equipartition magnetic field strength inside the regions of the radio emission as seen by LOFAR. To do this we used a slightly modified version of the equation given in \cite{2005AN....326..414B}:
\begin{equation}
    B_\textrm{eq} = \left( \frac{4\pi(1 -2\alpha)I_{\nu}E_i^{1+2\alpha}\left(\frac{\nu}{2c_1}\right)^{-\alpha}}{-(2\alpha + 1)c_2lc_4} \right)^{\frac{1}{(3 -\alpha)}} .
    \end{equation}
Here $\alpha$ is the spectral index of the regarded region, $I_\nu$ is the intensity of the synchrotron emission at the frequency $\nu$, $E_i$ is the lower energy cut due to ionisation losses and $l$ is the pathlength along the line of sight. The factors $c_1$, $c_2$ and $c_4$ are only dependant on the spectral index and are defined (with different spectral index convention) in \cite{2005AN....326..414B}.

This equation holds only under a number of assumptions. We follow the standard assumption that all cosmic rays are accelerated by electromagnetic processes and that there is a strong coupling between them and the magnetic fields. Additionally we assume that the relativistic plasma of the jets consists of only electrons and positrons without any contribution of protons. The formula we used is only valid for spectral indices with $\alpha \leq -0.5 $, otherwise the presence of ionisation or absorption processes may falsify the results. Hence, the core regions for each source have been excluded for this analysis. The lower energy cut was set to $\gamma_{\text{min}} = 100$ based on the derivations in \cite{1995A&A...293..665F}. To determine the pathlength $l$ we assumed a cylindrical shape of the radio emission. We derive magnetic field strengths of $2 - 10 \,\mu \mathrm{G}$ that are collected in Table \ref{tab:properties}.

In the work of \cite{2008ApJ...686..859B} energetics and particle content of the lobes of 24 radio galaxies has been analysed, which includes the calculation of the equipartition magnetic field strength for two sources which overlap with our sample. The values calculated for NGC\,4374 and NGC\,4486 are approximately twice as large as the magnetic field strengths we determined. This may be partly due to the large uncertainties in the calculation of the field strength, as it is based on assumptions for the volume of the radio emission and the particle content in the lobes. 
Additionally, \cite{2008ApJ...686..859B} use the 'classical' equipartition magnetic field strength first proposed by \cite{1956ApJ...124..416B}. There, the dependence of the integration limits over the synchrotron spectrum on the magnetic field was not taken into account, which can lead to an overestimation of fields with steep spectral index. This could be an explanation for the variation in our results.

\subsection{The wide-angle tail of NGC\,4472}
\label{subsec:ngc4472}
NGC\,4472 is the most massive and optically brightest galaxy in the Virgo cluster. In previous studies at Gigahertz-frequencies, this source displayed compact, double-lobed radio emission. Recently, \cite{2023A&A...676A..24E} reported the presence of extended radio tails at $\SI{144}{\mega\Hz}$. In this section, we will analyse the newly found radio tails and determine their interplay with the surrounding ICM.

Each of the two radio tails has a size of about $\SI{150}{kpc}$ in the LOFAR observations. However, most of this extended radio emission cannot be seen in observations at higher frequencies as is the case for the MeerKAT observations at $\SI{1.28}{\giga\Hz}$. To reveal more of the potential diffuse low-surface brightness emission in the MeerKAT image, we perform a source subtraction and low-resolution imaging procedure as explained in Section \ref{subsec:source-sub} which put a higher emphasis on the large scale emission of NGC\,4472. As indicated in the spectral index map in Figure \ref{fig:spidx_maps} the size of the visible radio tails in MeerKAT is still smaller than in LOFAR. We are therefore only able to determine an upper limit for the spectral index in the wide angle tails of $\alpha \leq -1.8$. As mentioned in Section \ref{subsec:specidx}, this implies that the wide-angle tail is from an earlier phase of activity of the central AGN, leaving behind spectral steepened, old radio emission.

Looking at the eastern and western radio tail in detail, they show different morphological behaviour. Especially the eastern tail shows interesting features, marked in Figure \ref{fig:radio_cav} by a dashed blue box. In this part the radio emission forms a narrow line which seems to be slightly detached from the emission closer to the core of the galaxy. This may possibly hint towards a rather confined magnetic field structure. In the western radio tail no similar features can be seen. The radio emission seems smooth and uninterrupted. This might indicate, together with the analysis of cavities and the ICM below, that the radio emission in the eastern arm is more affected by differing processes as the buoyant rise of radio bubbles and the ram pressure stripping, while the western arm seems to be mainly affected by ram pressure stripping. The ram pressure stripped X-ray tail of NGC\,4472 has been analysed previously in various X-ray studies for instance by \cite{2004ApJ...613..238B}, \cite{2011ApJ...727...41K} and \cite{2019AJ....158....6S}.

\begin{figure}[h]
     \centering
     \begin{subfigure}[h]{0.95\columnwidth}
        \centering
        \includegraphics[width=\linewidth]{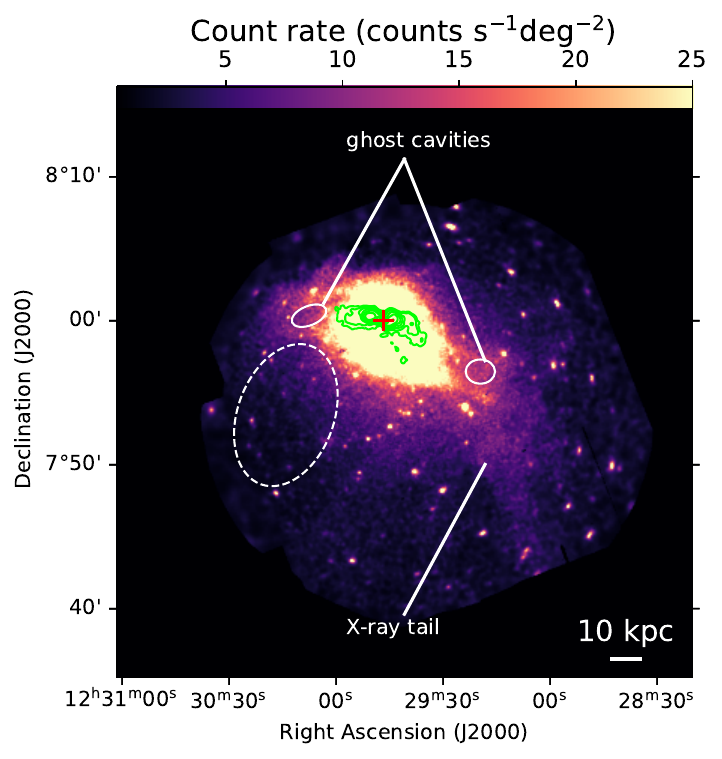}
    \end{subfigure}
    \begin{subfigure}[h]{0.95\columnwidth}
        \centering
        \includegraphics[width=\linewidth]{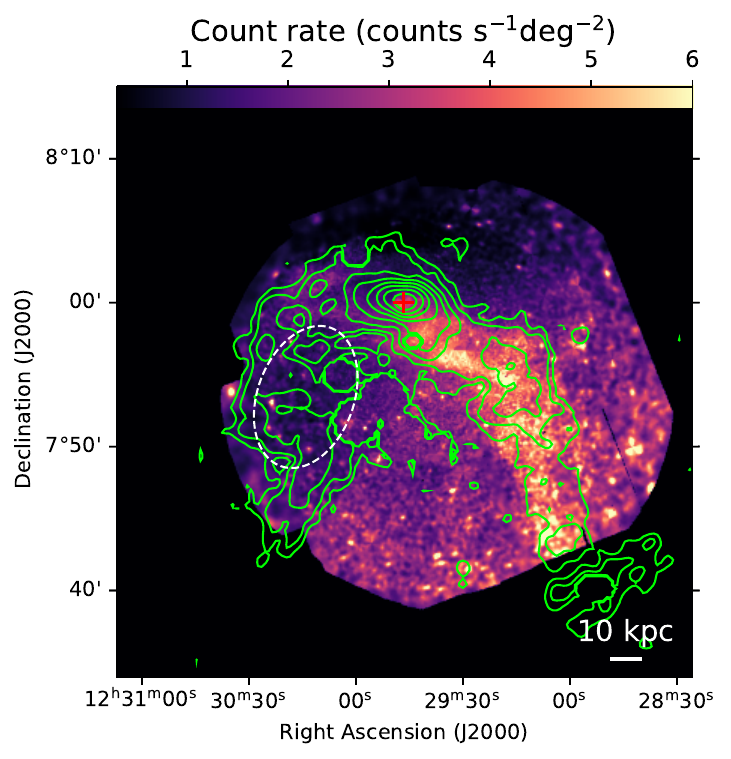}
    \end{subfigure}
     \caption{MeerKAT contours on XMM-Newton X-ray image of NGC\,4472 (top) and 1$'$ LOFAR contours on the $\beta$-model-subtracted X-ray image (bottom). Contours start at 3$\sigma$ and increase in powers of 2, where $\sigma{=}17\,\mu \mathrm{Jy}/\mathrm{beam}$ for MeerKAT and 1$\sigma{=}1 \,\mathrm{mJy}/\mathrm{beam}$ for LOFAR. The centre of Ngc\,4472 is marked with a red cross.}
     \label{fig:ngc4472_xray}
\end{figure}
To gain a better understanding of the behaviour of NGC\,4472 and to analyse the influence and effect of the surrounding ICM we overlay the radio contours as seen with LOFAR and MeerKAT on the $\beta$-model subtracted residual X-ray image (see Section \ref{subsec:xmm-newton}). This is shown in Figure \ref{fig:ngc4472_xray}. It is clearly visible that the western X-ray extension coincides strongly with the extended radio emission seen in that part. NGC\,4472 seems to be falling onto the Virgo cluster almost perpendicular to the line of sight \citep{2007ApJ...655..144M}. This infall likely causes the ram pressure stripping of the gas forming the extended X-ray tail seen in Figure \ref{fig:ngc4472_xray}. This has been analysed in-depth by \cite{2019AJ....158....6S}. The same effect caused the bent shape of the radio jets \citep{2023A&A...676A..24E}. With the radio and X-ray tail overlapping over the full extent of the tail, it is very likely that the X-ray tail is strongly connected to the AGN outburst. This result supports the explanation by \cite{2019AJ....158....6S} that an AGN outburst could have displaced gas from the ISM to larger radii, facilitating the stripping process.

Surprisingly we do not see a similar behaviour in the eastern radio lobe of the source. Despite the bent shape of the radio jet as it is the case in the western tail, we do not see enhanced X-ray emission. Instead we see a depression of surface brightness with its centre at a distance of $\sim \SI{24}{\kilo\parsec}$ from the galaxy centre. A comparison of the count rate in this area with the surrounding space alone does not yield a statistically significant lack of X-ray emission, with the values being within $2\sigma$ of each other. However, the shape of the surface brightness depression follows the shape of the radio tail closely, strongly implying that this feature is indeed an X-ray cavity. We, therefore measure the size of this cavity and determine an estimate for the jet power.

To determine the volume, we assumed an ellipsoidal shape of the cavity with its size along the line of sight the same as the semi-minor axis of the ellipse fitted to it. To calculate the enthalpy required to inflate the cavity given by $E_\mathrm{cav} = 4pV$, we used the deprojected pressure of the ICM close to the cavity. The pressure was calculated using the temperature and density profiles derived from a region between the two radio arms as explained in Section \ref{subsec:xmm-newton}. We estimated the age of the cavity using the buoyancy timescale as given by \cite{2004ApJ...607..800B}:
\begin{equation}
    t_{\mathrm{buoy}} = R \sqrt{\frac{SC}{2gV}} ,
\end{equation}
where R is the projected radial distance to the centre of NGC\,4472, S is the cross-sectional area of the cavity, C is the drag coefficient chosen to be $C = 0.75$ \citep{2001ApJ...554..261C}, V is the volume of the cavity and g is the gravitational acceleration. We follow \cite{2004ApJ...607..800B} and approximate the gravitational acceleration with $g \approx 2v_{\sigma}^2/R$ with the stellar velocity dispersion of NGC\,4472 given as $v_{\sigma} = \SI{312}{\km\per\s}$ \citep{2003ApJ...591..850C}. We find an age of $t_{\mathrm{buoy}} \approx \SI{83}{\mega\years}$. We computed the cavity power for this cavity to be $P_{\mathrm{cav}} = \SI{2.18e42}{\erg\per\s}$.

In comparison to the age of the cavity, the time it takes for the X-ray tail to form is estimated to be $100 - 220 \,$Myrs \citep{2019AJ....158....6S}. For an AGN outburst we would expect the radio tails to form on similar timescales. As the X-ray emission seen in NGC\,4472 seems to be correlated with the radio tails as well, we would expect the cavity to have a similar age as the X-ray tail. There are, however, multiple ways in which can explain this difference in timescales. The cavity age we derived in this paper is more likely to be a lower limit of the actual cavity age. For its estimation we used its radial distance to the centre of NGC\,4472. With the shape of the radio tails, however, it is more likely that the radio emission had a more complicated path towards its position which would result in an older age of the cavity. Additionally NGC\,4472 is falling into the Virgo cluster in the north-east direction and experiences ram pressure stripping towards the south-west \citep{2019AJ....158....6S}. This could affect the two jets of the central AGN differently in their evolution and timescales, with the western jet produced in the approximate direction of the stripping and the eastern jet more orthogonal to it. 

NGC\,4472 contains another set of cavities located at a distance of $\SI{3.6}{\kilo\parsec}$ from the galaxy centre. They have been found and analysed in the work of \cite{2004ApJ...613..238B}. There, they found that these cavities were produced by an AGN outburst that began at least $\SI{1.2e7}{yrs}$ ago and may still be active today. As the cavity we found was produced approximately $ \SI{83}{\mega\years}$ ago and due to the fact that the radio emission filling this cavity is only visible at very low frequencies, we believe that these cavities have been produced by two separate outbursts of the AGN. The estimated jet power of the cavities found by \cite{2004ApJ...613..238B} is $P_{\mathrm{cav}} \approx \SI{0.5e42}{\erg\per\s}$ \citep{2010ApJ...720.1066C}. This jet power is approximately as high as the jet power we estimated for the previous outburst of the AGN, implying that the two outbursts are similar in strength. 

\begin{figure}[h]
     \centering
     \includegraphics[width=0.95\columnwidth]{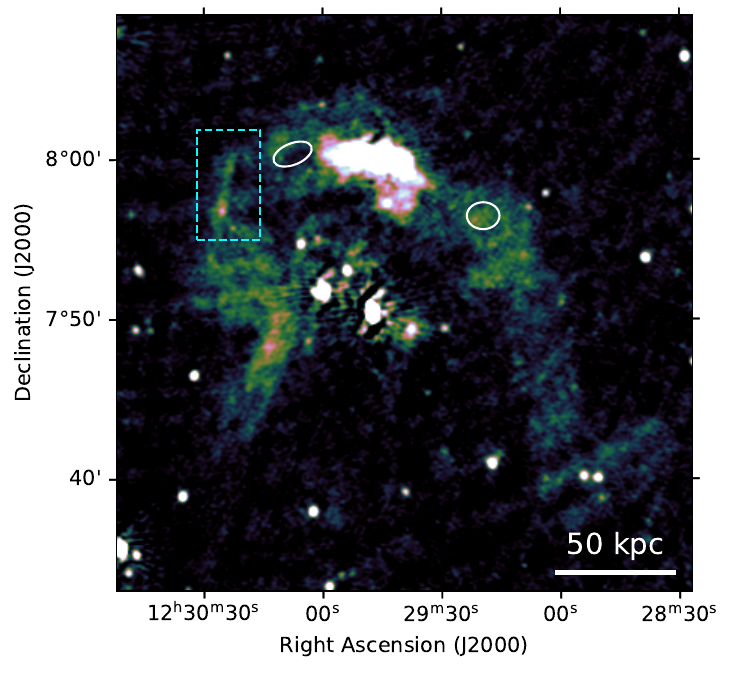}
     \caption{Cavities marked on a low resolution LOFAR image of NGC\,4472. The two cavities marked with white ellipses are the ghost cavities found by \cite{2011ApJ...727...41K}. The dashed blue box marks a part of collimated radio emission.}
     \label{fig:radio_cav}
\end{figure}
NGC\,4472 may harbour two other cavities at larger distance from the centre of the galaxy, as has been proposed by \cite{2011ApJ...727...41K}. In X-ray observations of the source they found cool filamentary arms, implying the existence of cavities from a previous outburst of the AGN. The presence of these filaments has been confirmed in a deeper X-ray study by \cite{2019AJ....158....6S} and can be seen in the lower image of Figure \ref{fig:ngc4472_xray}. For cavities produced by the buoyant rise of radio bubbles we would expect them to be filled with radio emission. This does not seem to be the case for the eastern cavity. Here it seems that on the contrary the radio emission declines in a region close to the cavity. This depression of radio emission is approximately of the same size as the cavity but slightly offset from the exact position. We therefore cannot conclusively determine whether this is a real cavity. The western ghost cavity on the other hand is radio-filled.   

\begin{figure}[h]
     \centering
     \includegraphics[width=0.9\columnwidth]{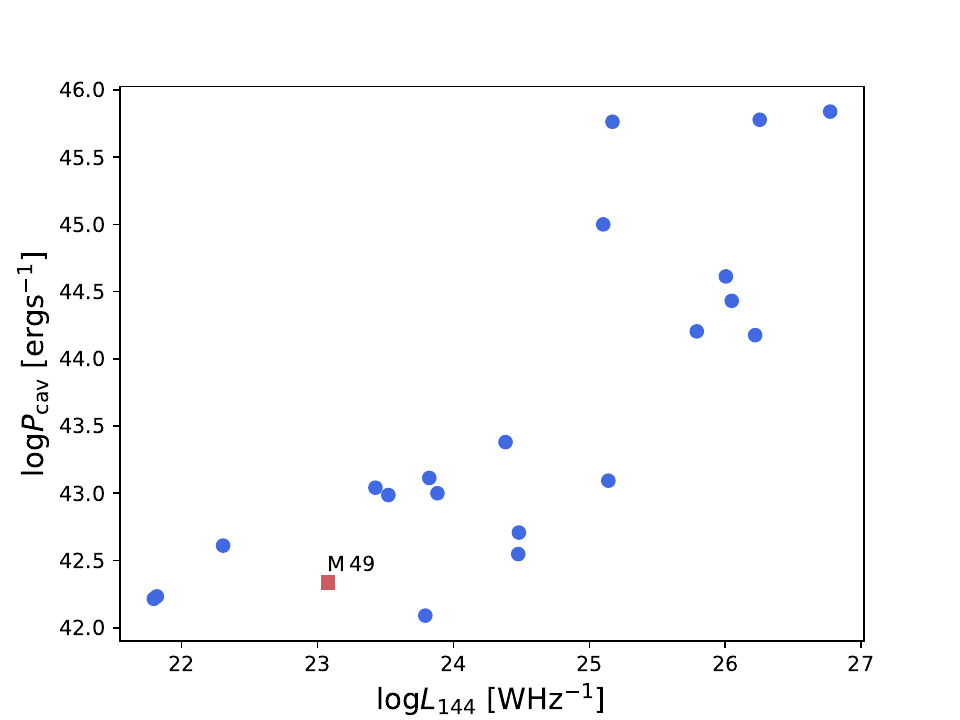}
     \caption{Cavity power against the total radio luminosity of each source. The plot was created using the data from \cite{2020MNRAS.496.2613B}. The point marked as a red square has been added using the cavity power estimated for NGC\,4472 (M\,49) in this work.}
     \label{fig:cav_plot}
\end{figure}

However, we use only the cavity power calculated for the single cavity found in this paper as an estimation of the total jet power of the AGN for the past duty cycle. To check whether our estimated jet power is reasonable, we determine its position in the jet power against radio luminosity plot. A scaling relation in this plot has been found in earlier work \citep[e.g. by][]{2004ApJ...607..800B,2008ApJ...686..859B,2010ApJ...720.1066C,2011ApJ...735...11O} at higher frequencies. Here, we use the data and sample assembled by \cite{2020MNRAS.496.2613B} to create a plot at 144\,MHz which can be seen in Figure \ref{fig:cav_plot}. The references for calculating these cavity powers can be found in \cite{2020MNRAS.496.2613B}
As the cavity is located at a greater distance from the centre of the galaxy, an age estimation of a buoyantly rising bubble is quite high. The jet power derived from this is therefore at the lower end of the compared to the jet power of other sources gathered in the sample. With the radio luminosity of NGC\,4472 being on the lower side as well, the cavity fits into the correlation. However, as the scatter in the plot is quite large, this is only a weak indication that we have found a real cavity.

\subsection{Remnant source NGC\,4365}
\label{subsec:N4365}

In this section we take a closer look at NGC\,4365, a giant early-type galaxy which is the dominant object in the $W^\prime$-cloud in the background of the Virgo cluster ($d=23.3\,$Mpc). Like the centres of other substructures in the Virgo cluster, it has likely been formed during a major merging event \citep{2014A&A...570A..69B}.

\begin{figure}[h]
     \centering
     \includegraphics[width=0.9\columnwidth]{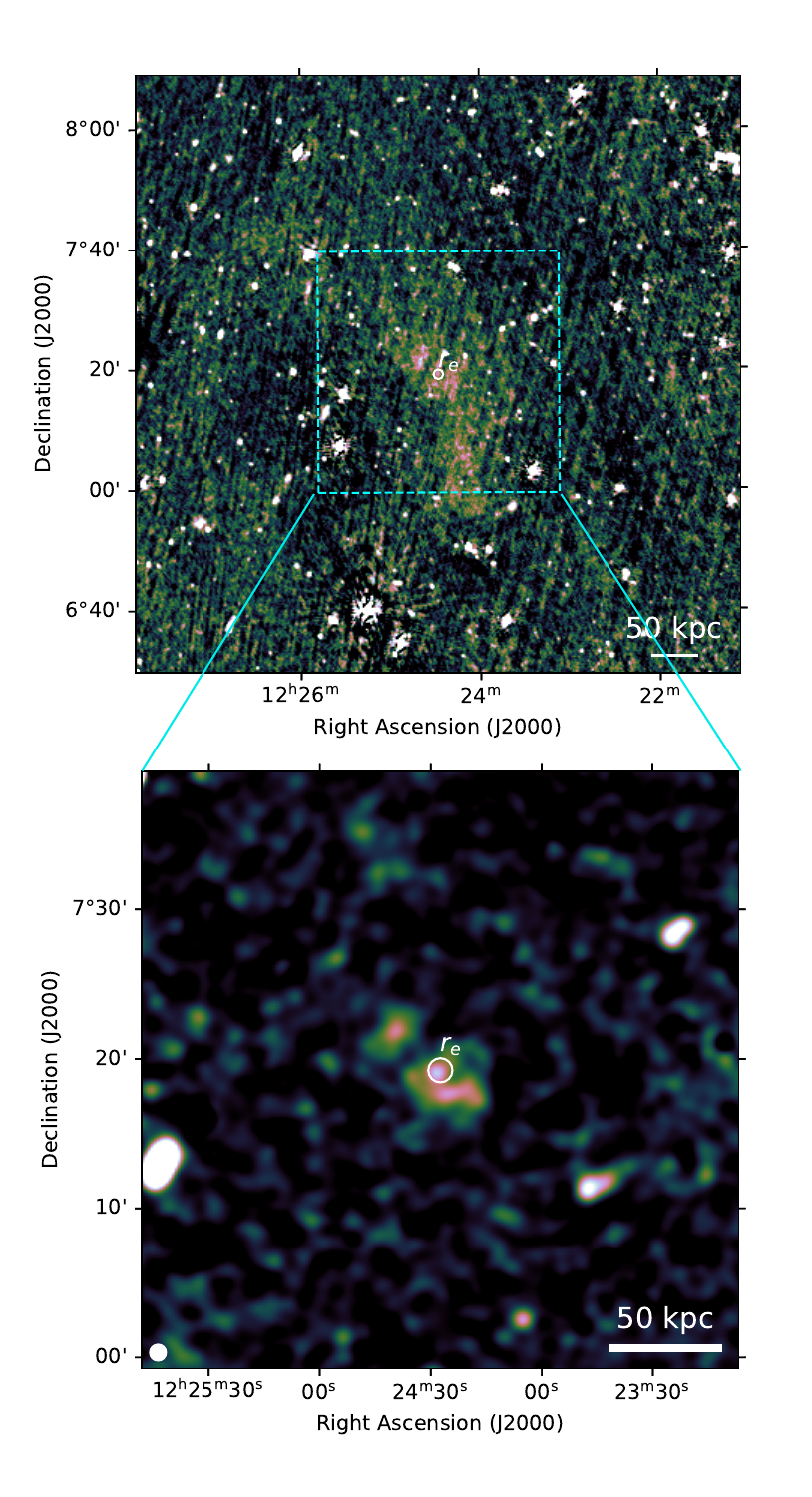}
     \caption{Low-resolution radio map of the diffuse emission surrounding NGC\,4365 at 144\,MHz (LOFAR, top panel). The bottom panel shows the area corresponding to the cyan box in the MeerKAT compact-source-subtracted  map at $60''$. The white circle in the middle of both images marks the half-light radius in the $r$-band of NGC\,4365.}
     \label{fig:ngc4365_radio}
\end{figure}

For this galaxy, we can rule out compact radio emission with a luminosity above $L_{144}=\SI{6.75e19}{\W\per\Hz}$, meaning that this galaxy is surprisingly passive for its mass \cite[compare e.g.][]{2019A&A...622A..17S}. In the low resolution LOFAR images we can instead see very diffuse radio emission extending over approximately $\SI{3790}{arcsec}$ which corresponds to $\SI{428}{\kilo\parsec}$. This emission is surrounding the optical position of NGC\,4365, as has been found by \cite{2023A&A...676A..24E} and as can be seen in Figure \ref{fig:ngc4365_radio}.

The radio emission seen with LOFAR is not necessarily connected to NGC\,4365. In eROSITA observations \cite{2024arXiv240117296M} found X-ray emission extending from the NGC\,4472 group over the W' cloud. They argue that based on the coincident position and morphology the radio emission is connected to the X-ray and caused by accretion shocks or turbulence in the hot plasma \citep{2019SSRv..215...16V}.

Hints on whether the radio emission stems from NGC\,4365 or from other causes can be found in the MeerKAT observations. As the radio emission is very diffuse in the LOFAR images we expect any possible extended radio emission to be diffuse as well in the MeerKAT images. We therefore created a source-subtracted radio image to highlight faint extended emission as explained in Section \ref{subsec:source-sub}. This image is shown in Figure \ref{fig:ngc4365_radio}. Here, we can see a faint point source at the optical position of NGC\,4365 which sits at the centre of diffuse radio emission surrounding it. Its extension of $\approx70$\,kpc is much less  than the radio emission seen with LOFAR with a size of $\sim600''$.

We conclude that  the emission seen at the two different frequencies is connected to each other and originate from a past AGN outburst of NGC\,4365. The missing point source in the LOFAR observations of NGC\,4365 could be very faint and not differentiable from the extended emission. In that case, the X-ray extension found by \cite{2024arXiv240117296M}, may not be connected to the extended radio emission seen with LOFAR and coincidentally shows the same orientation.

\begin{figure}[h]
     \centering
     \includegraphics[width=0.9\columnwidth]{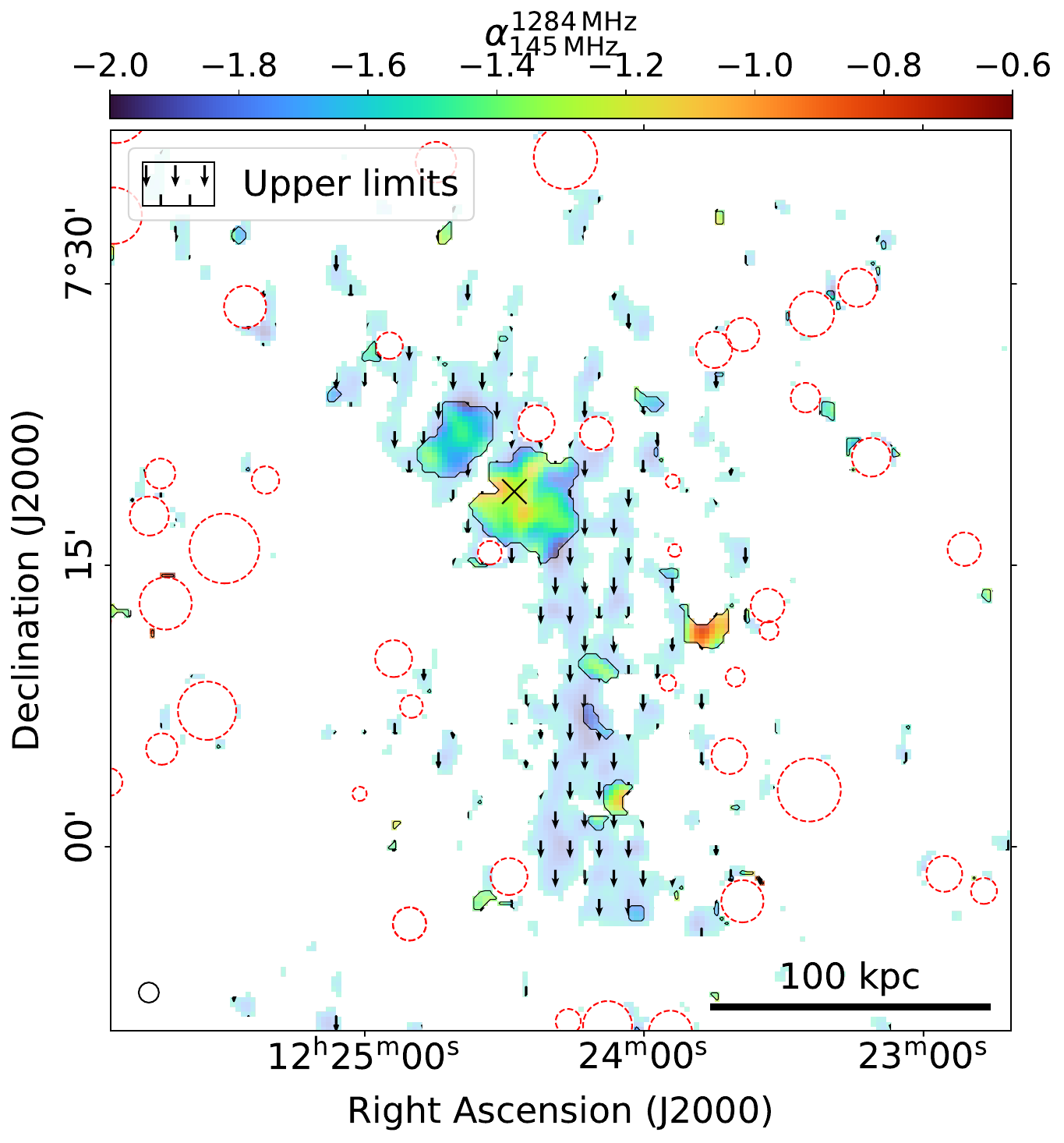}
     \caption{Spectral index map of NGC\,4365 (black cross). Dashed red circles mark unrelated sources that were masked. Upper limits are marked by downward pointing arrows, the synthesized beam size and linear size at the distance of the source are marked at the bottom.}
     \label{fig:spidx_4365}
\end{figure}

Based on the compact-source subtracted MeerKAT and the compact source-masked LOFAR maps at 1$'$ resolution, we create a spectral index map of the extended emission. For the much larger extent of the radio emission in LOFAR we can estimate an upper limit of the spectral index. This is shown in Figure \ref{fig:spidx_4365}, the corresponding uncertainty map can be found in \autoref{sec:app_spidxerr}.

In the map we can see that the source has generally a very steep spectrum. The spectral index is slightly flatter in the centre with a value of $\alpha \sim -1.1$. At larger distance from the galaxy, the emission becomes steeper until it is only detected in LOFAR and we can only determine an upper limit of $\alpha < -1.6$.  This is in agreement with remnant AGN emission from a previous AGN-outburst with a spectral index gradient pointing towards older plasma further away from NGC\,4365. Thus, while the galaxy appears to be particularly passive at present day (see \autoref{fig:mass_lum}), it likely was significantly more active in the past and is now in the remnant AGN stage.

\section{Conclusions}
\label{sec:conclusion}

In this work we analysed the radio properties of a sample of 94 early-type galaxies in the Virgo cluster using LOFAR data at $\SI{144}{\MHz}$. Twelve of these galaxies detected at a significance of $4\sigma$ in LOFAR. We determined the relations between the radio luminosity, mass of the host galaxy and largest linear size of the emission. Data from MeerKAT L-band observations was available for eleven galaxies in the sample, making an analysis of the spectral properties possible. 
The main results for the statistical analysis of the sample are:

\begin{itemize}
    \item We find a correlation between the observed radio luminosity and the stellar mass of the host galaxy. This is consistent with the results in \cite{2019A&A...622A..17S} and the expectation from relations between the mass and fraction of radio-observed galaxies \citep[e.g. ][]{2005MNRAS.362...25B,2019A&A...622A..17S}.
    \item We can confirm a positive connection between the radio luminosity and linear size of the sources in agreement with \cite{2022A&A...660A..93C} 
    \item We found hints for a possible correlation between the spectral index and the linear size of the sources. The larger sources show steeper radio continuum spectra. We speculate that this may only be visible for extended sources in similar environments.
\end{itemize}

Special consideration was given to NGC\,4472 which was analysed regarding the interaction between the radio lobes and the X-ray emitting ICM. 
Furthermore, we investigated in more detail the galaxy NGC\,4365, which was not detected at a 4$\sigma$ level in LOFAR observations, making it particularly radio-passive for a galaxy of its mass. 
The results for this study on individual galaxies are: 

\begin{itemize}
    \item The co-evolution of the thermal plasma and the cosmic ray electrons in the eastern and western tail of NGC\,4472 is drastically different. While the western tail is dominated by ram pressure stripping
    \citep[see also]{2019AJ....158....6S}, the eastern tail hints towards a cavity that is inflated by the radio emission. The cavity power is $P_{\mathrm{cav}} = \SI{2.18e42}{\erg\per\s}$, which sits at the lower end in a cavity power - radio luminosity correlation.
    \item Comparing two different duty cycles in NGC\,4472 based on the jet powers of two outbursts yields that the past outburst is of the same order as the present one.
    \item We cannot confirm the existence of previously found eastern ghost cavity in NGC\,4472 with our radio data. It rather seems that the radio emission declines at its position.
    \item A notable outlier from the stellar mass-luminosity relation is NGC\,4365, which does not exhibit significant compact radio emission in spite of its high mass.
    \item NGC\,4365 is surrounded by diffuse, low-surface brightness emission at $\SI{144}{\mega\hertz}$ with an extent of $\sim 1$ degree (410\,kpc at the distance of NGC\,4365). This emission is only partly visible in MeerKAT after low-resolution imaging which suggests that this emission is steep-spectrum (with $\alpha{<}-1.6$ in the extended tails seen only with LOFAR). We interpret it as fossil radio galaxy plasma from a past AGN outburst.
\end{itemize}

These results are based on the analysis of a relatively small set of sources. A larger sample needs to be studied to strengthen the statistical result. All the sources analysed are part of sub-structures belonging to the same galaxy cluster. Therefore, some of the connections found here may be more challenging to recover in a sample of sources in more diverse astrophysical surroundings. A more conclusive picture, especially in regards to the possible connection between the spectral index and the linear size, could be achieved by analysing a larger sample of AGN over various clusters and in comparison to field galaxies.

For the extended steep-spectrum emission of NGC\,4472 and NGC\,4365, studies at even lower frequencies are important to establish a more coherent picture, in particular to model the spectral ageing. The observations of our ongoing ViCTORIA LOFAR LBA survey at frequencies of $42-66$~MHz will allow us to confirm our interpretation of the radio emission. For both galaxies, the LOFAR LBA observations may reveal even older and even more distant radio plasma. The extended radio tails of NGC\,4472 are at their furthest point not visible in the MeerKAT observations. Therefore only upper limits for the spectral index can be estimated in these regions. With the LBA data it is possible to determine spectral index maps for the tails and to do a proper spectral analysis, which sheds more light on the properties of this past outburst.
LBA observations may also potentially be decisive for the question whether the diffuse radio emission is connected to NGC\,4365 or has another origin.

For NGC\,4472 the structure of the radio emission in the eastern arm poses an additional research topic. In the discussion above we speculated that the narrow jet-like form of the radio emission may arise from a structured magnetic filed. This idea can be confirmed or disproved with a polarisation analysis of NGC\,4472. 

\begin{acknowledgements}
HE acknowledges support by the Deutsche Forschungsgemeinschaft (DFG, German Research Foundation) under project number 427771150. MB acknowledges funding by the Deutsche Forschungsgemeinschaft (DFG, German Research Foundation) under Germany's Excellence Strategy -- EXC 2121 ``Quantum Universe'' --  390833306 and project number 443220636 (DFG research unit FOR 5195: "Relativistic Jets in Active Galaxies").

FdG acknowledges support from the ERC Consolidator Grant ULU 101086378.

The MeerKAT telescope is operated by the South African Radio Astronomy Observatory, which is a facility
of the National Research Foundation, an agency of the Department of Science and Innovation.
Part of the data published here have been reduced using the CARACal pipeline, partially supported by ERC Starting grant number 679627 “FORNAX”, MAECI Grant Number ZA18GR02, DST-NRF Grant Number 113121 as part of the ISARP Joint Research Scheme, and BMBF project 05A17PC2 for D-MeerKAT. Information about CARACal can be obtained online under the URL: \url{https://caracal.readthedocs.io}.

LOFAR (van Haarlem et al. 2013) is the Low Frequency Array designed and constructed by
ASTRON. It has observing, data processing, and data storage facilities in several countries, which are owned by various parties (each with their own funding sources), and that are collectively operated by the ILT foundation under a joint scientific policy. The ILT resources have benefited from the following recent major funding sources: CNRS-INSU, Observatoire de Paris and Université d'Orléans, France; BMBF, MIWF-NRW, MPG, Germany; Science Foundation Ireland (SFI), Department of Business, Enterprise and Innovation (DBEI), Ireland; NWO, The Netherlands; The Science and Technology Facilities Council, UK; Ministry of Science and Higher Education, Poland; The Istituto Nazionale di Astrofisica (INAF), Italy. This research made use of the Dutch national e-infrastructure with support of the SURF Cooperative (e-infra 180169) and the LOFAR e-infra group. The Jülich LOFAR Long Term Archive and the German LOFAR network are both coordinated and operated by the Jülich Supercomputing Centre (JSC), and computing resources on the supercomputer JUWELS at JSC were provided by the Gauss Centre for Supercomputing e.V. (grant CHTB00) through the John von Neumann Institute for Computing (NIC).
This research made use of the University of Hertfordshire high-performance computing facility and the LOFAR-UK computing facility located at the University of Hertfordshire and supported by STFC [ST/P000096/1], and of the Italian LOFAR IT computing infrastructure supported and operated by INAF, and by the Physics Department of Turin university (under an agreement with Consorzio Interuniversitario per la Fisica Spaziale) at the C3S Supercomputing Centre, Italy.
\end{acknowledgements}

\bibliographystyle{aa}
\bibliography{bibfile}

%
%

\onecolumn
\begin{appendix}

\section{LOFAR images of sample sources}
\label{sec:app_lofar}
\begin{figure*}[h]
     \centering
     \begin{subfigure}[h]{0.26\textwidth}
     \centering
     \includegraphics[width=\textwidth]{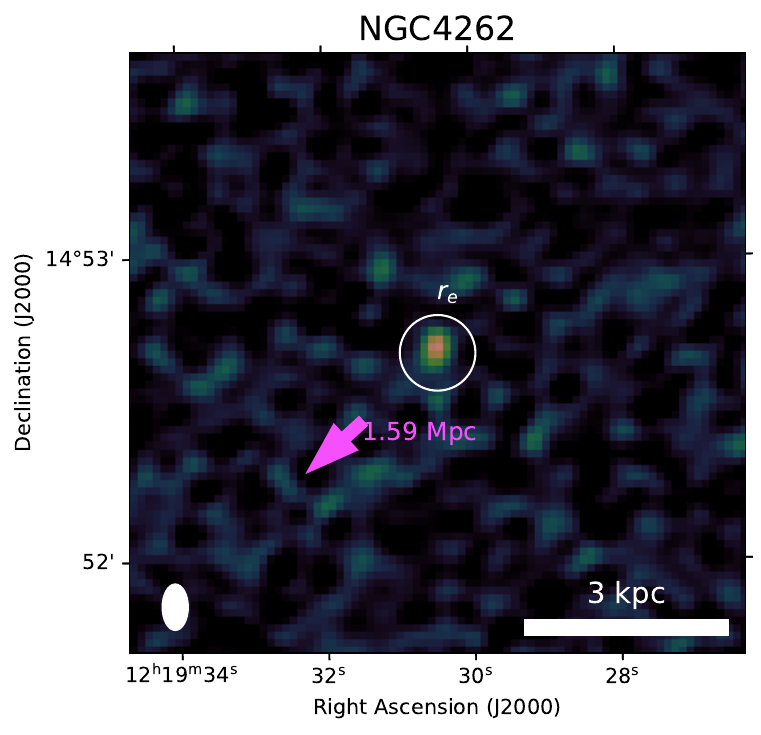}
     \label{fig:ngc4262_overlay}
     \end{subfigure}
     \hfill
     \begin{subfigure}[h]{0.26\textwidth}
         \centering
         \includegraphics[width=\textwidth]{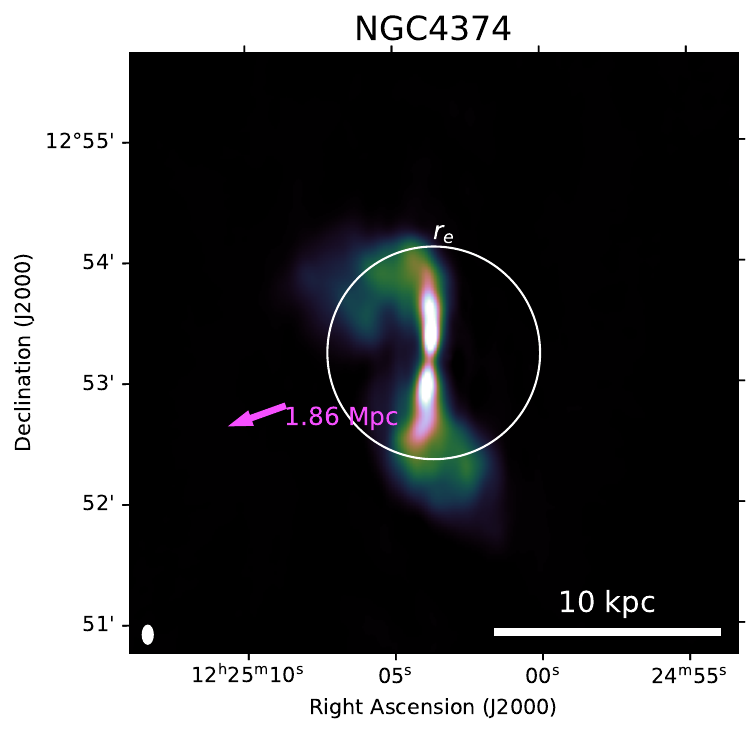}
         \label{fig:ngc4374_overlay}
     \end{subfigure}
     \hfill
     \begin{subfigure}[h]{0.26\textwidth}
         \centering
         \includegraphics[width=\textwidth]{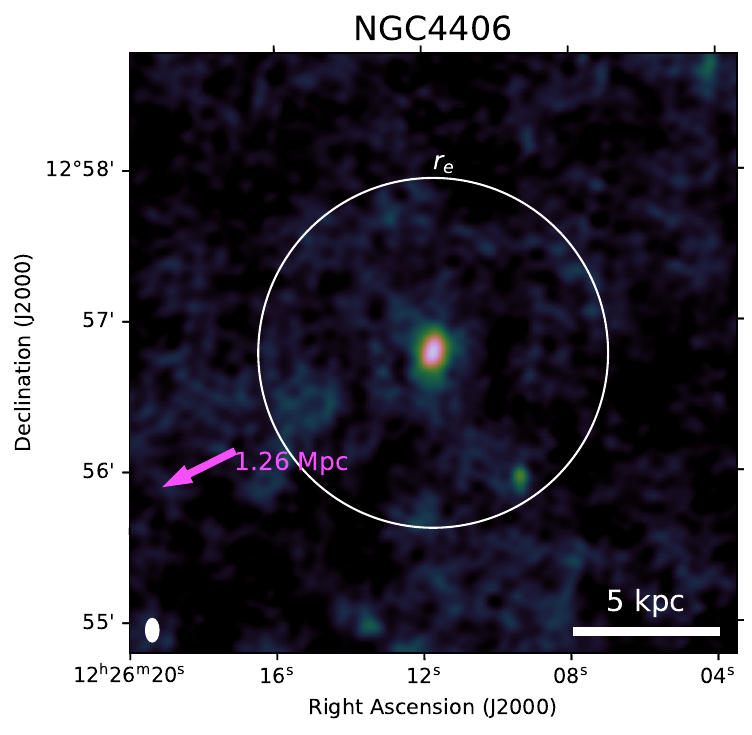}
         \label{fig:ngc4406_overlay}
     \end{subfigure}
     \begin{subfigure}[h]{0.26\textwidth}
         \centering
         \includegraphics[width=\textwidth]{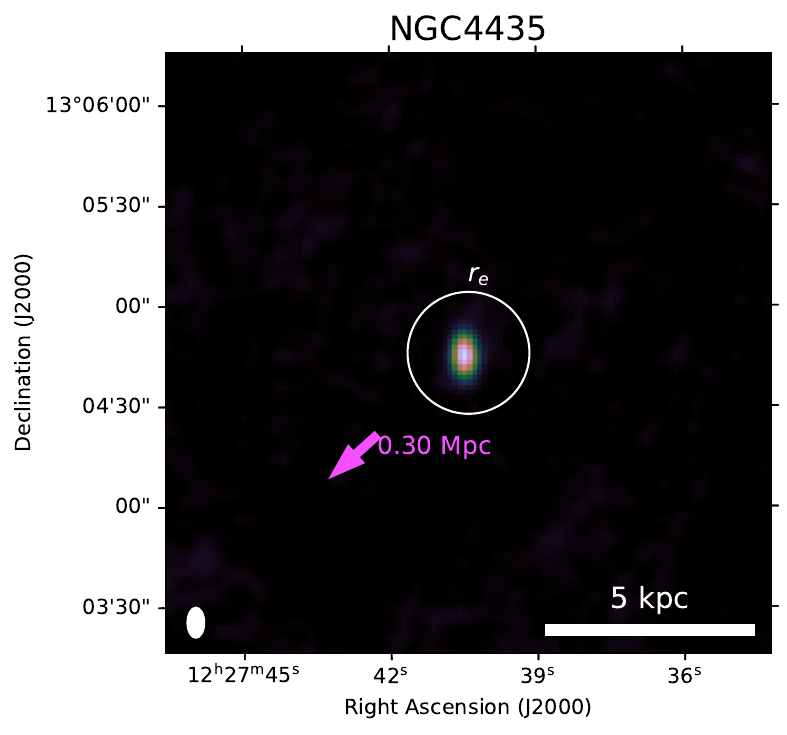}
         \label{fig:ngc4435_overlay}
     \end{subfigure}
     \hfill
     \begin{subfigure}[h]{0.26\textwidth}
         \centering
         \includegraphics[width=\textwidth]{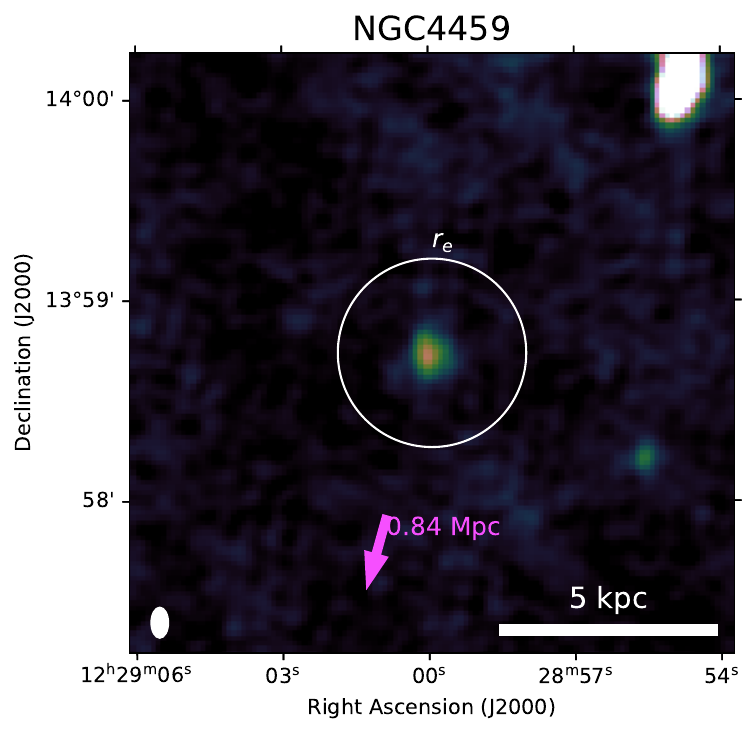}
         \label{fig:ngc4459_overlay}
     \end{subfigure}
     \hfill
     \begin{subfigure}[h]{0.26\textwidth}
         \centering
         \includegraphics[width=\textwidth]{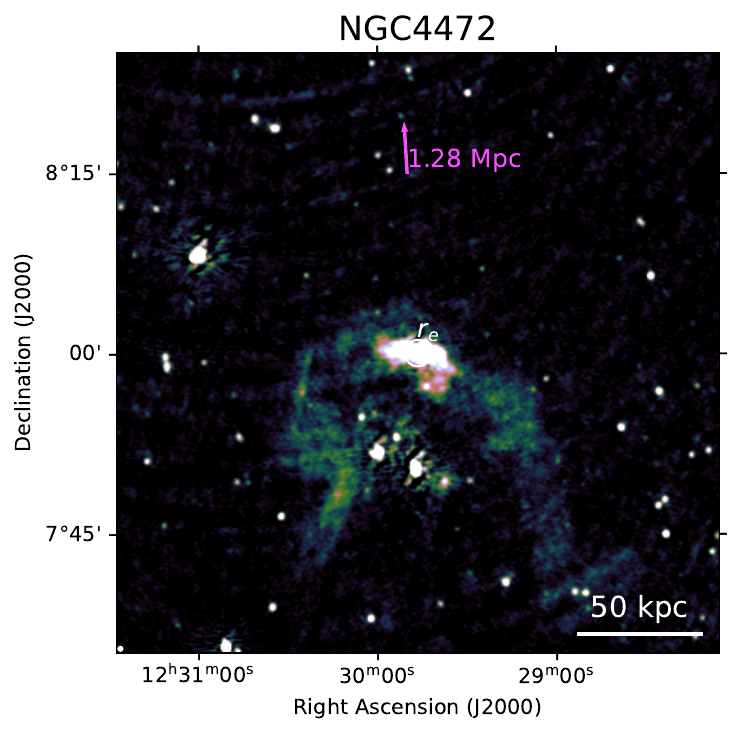}
         \label{fig:ngc4472_overlay}
     \end{subfigure}
     \begin{subfigure}[h]{0.26\textwidth}
         \centering
         \includegraphics[width=\textwidth]{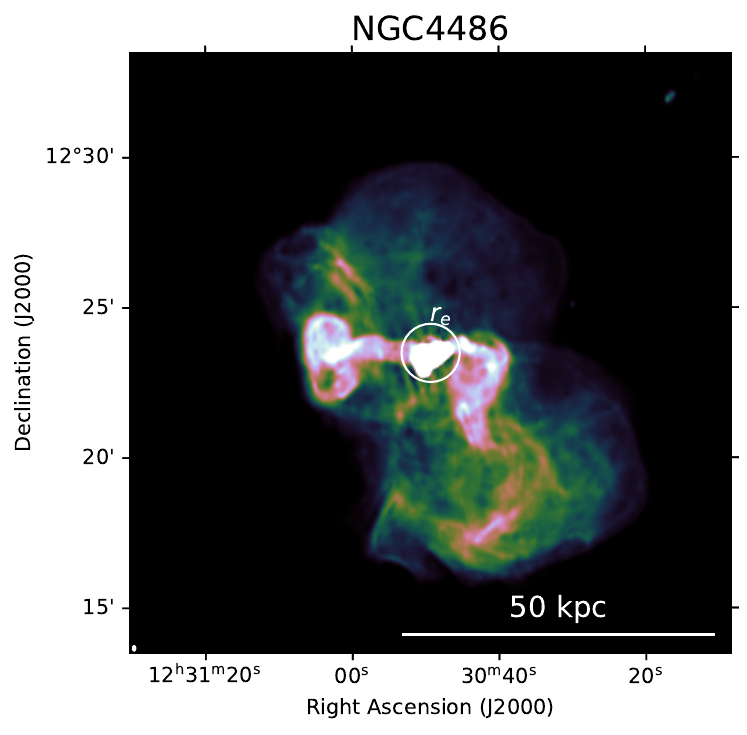}
         \label{fig:ngc4486_overlay}
     \end{subfigure}
     \hfill
     \begin{subfigure}[h]{0.26\textwidth}
         \centering
         \includegraphics[width=\textwidth]{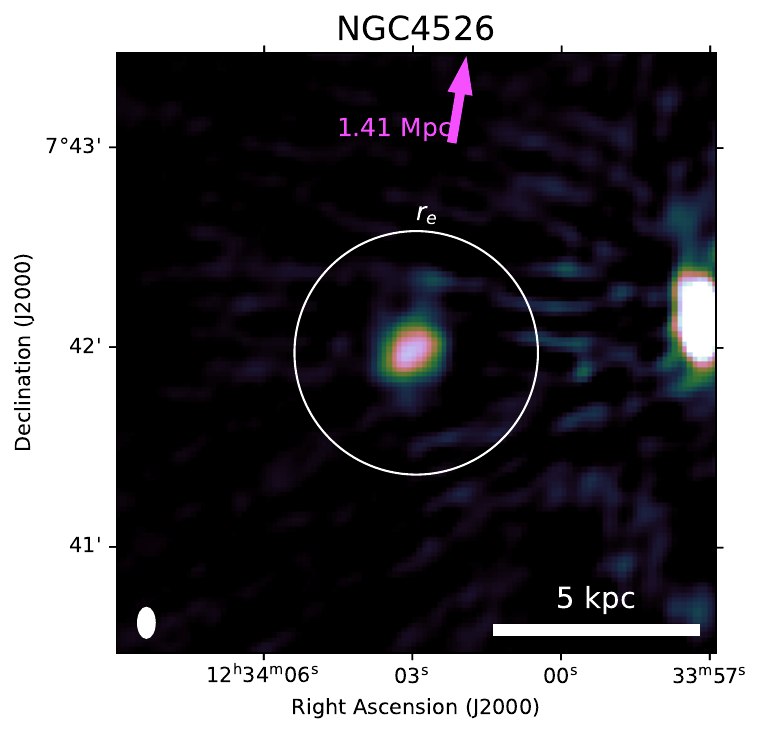}
         \label{fig:ngc4526_overlay}
     \end{subfigure}
     \hfill
     \begin{subfigure}[h]{0.26\textwidth}
         \centering
         \includegraphics[width=\textwidth]{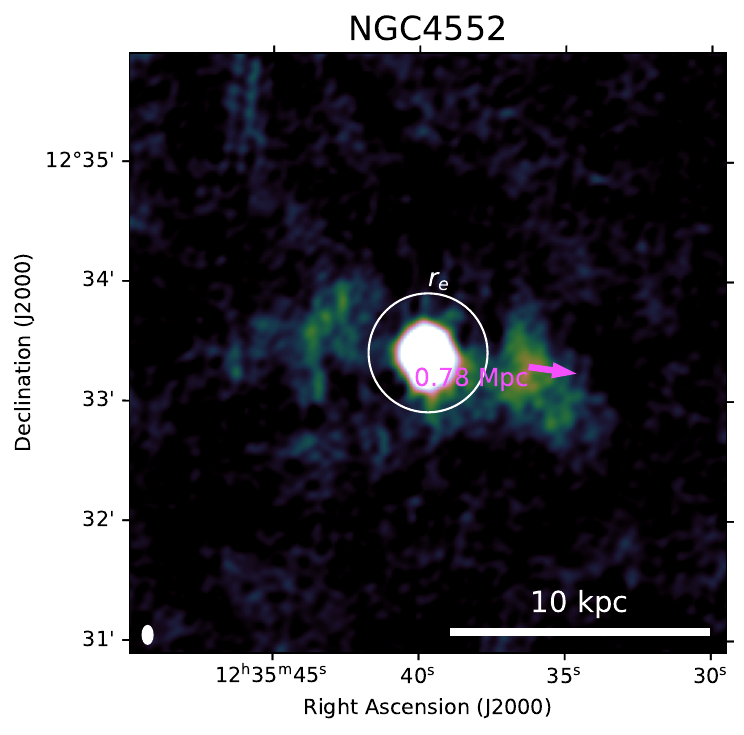}
         \label{fig:ngc4552_overlay}
     \end{subfigure}
     \begin{subfigure}[h]{0.26\textwidth}
         \centering
         \includegraphics[width=\textwidth]{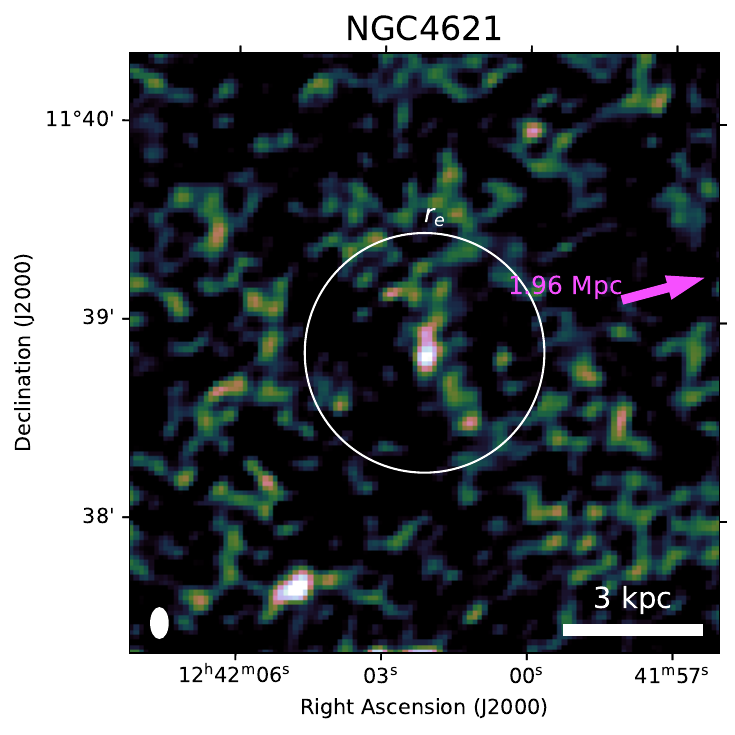}
         \label{fig:ngc4621_overlay}
     \end{subfigure}
     \hfill
     \begin{subfigure}[h]{0.26\textwidth}
         \centering
         \includegraphics[width=\textwidth]{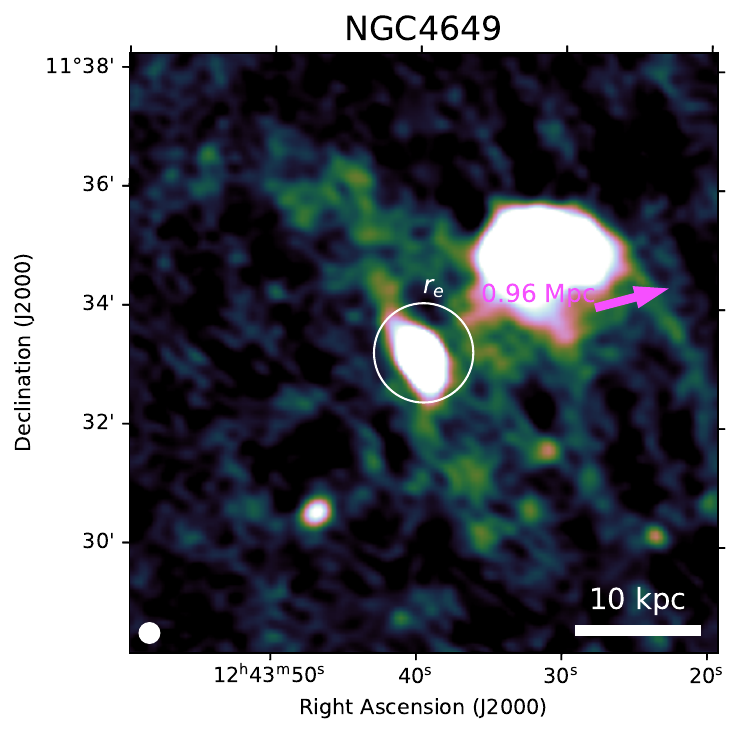}
         \label{fig:ngc4649_overlay}
     \end{subfigure}
     \hfill
     \begin{subfigure}[h]{0.26\textwidth}
         \centering
         \includegraphics[width=\textwidth]{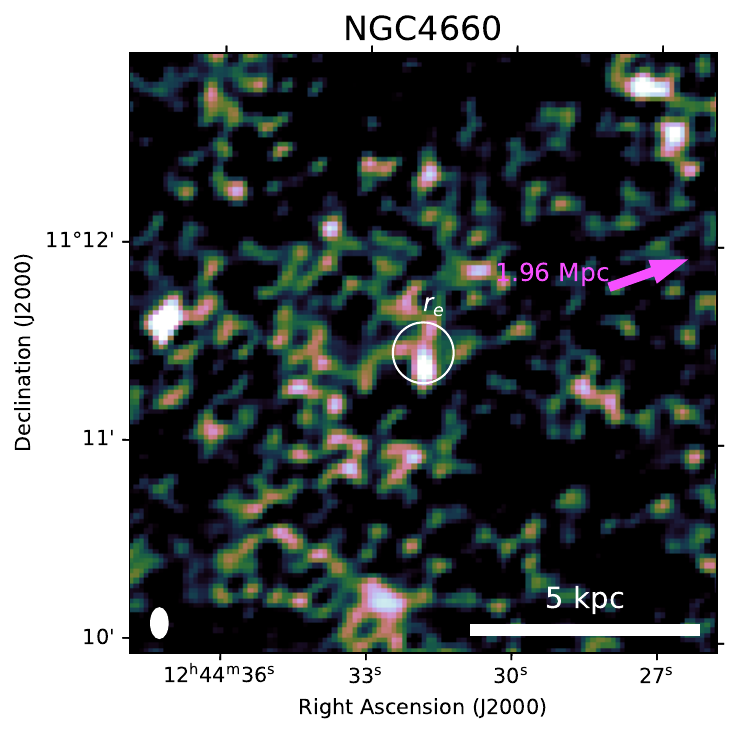}
         \label{fig:ngc4660_overlay}
     \end{subfigure}
    \caption{LOFAR image of the 12 sources in the sample at \SI{144}{MHz}. Each source is overlayed with the half light radius in the r-band of its host galaxy (white circle) taken from \cite{2014ApJS..215...22K} The pink arrow marks the direction and physical distance towards the centre of the Virgo cluster (NGC\,4486). The ellipse in the lower left corner indicates the beam size $9''\times5''$ for all images with the exception of NGC\,4472 and NGC\,4649. Here, the ellipse indicates a beam size of $20''\times20''$ as the low resolution LOFAR images are plotted.}
    \label{fig:radio_overlays}
\end{figure*}

\section{MeerKAT images of sample sources}
\label{sec:app_meerkat}
\begin{figure*}[h]
     \centering
     \begin{subfigure}[h]{0.26\textwidth}
     \centering
     \includegraphics[width=\textwidth]{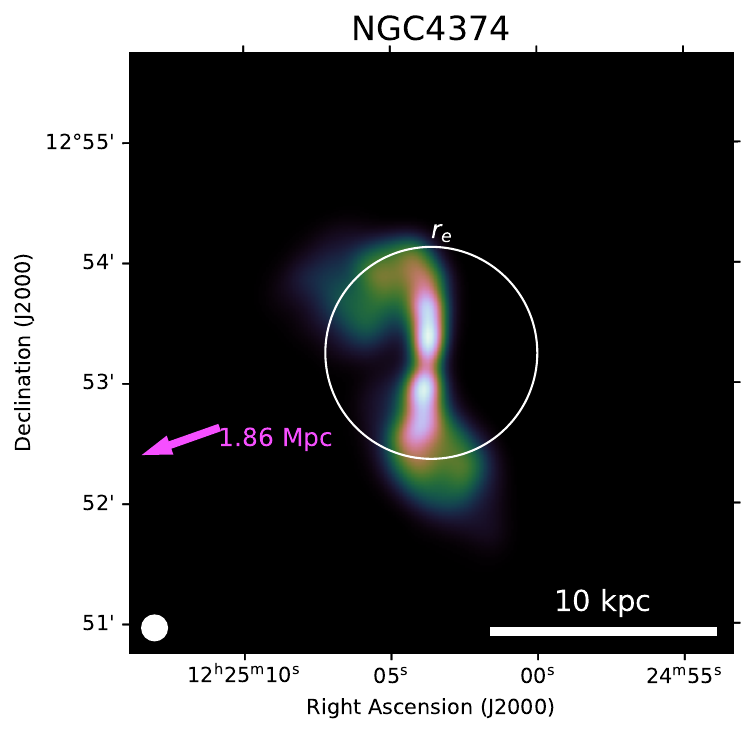}
     \label{fig:ngc4374_meerkat}
     \end{subfigure}
     \hfill
     \begin{subfigure}[h]{0.26\textwidth}
         \centering
         \includegraphics[width=\textwidth]{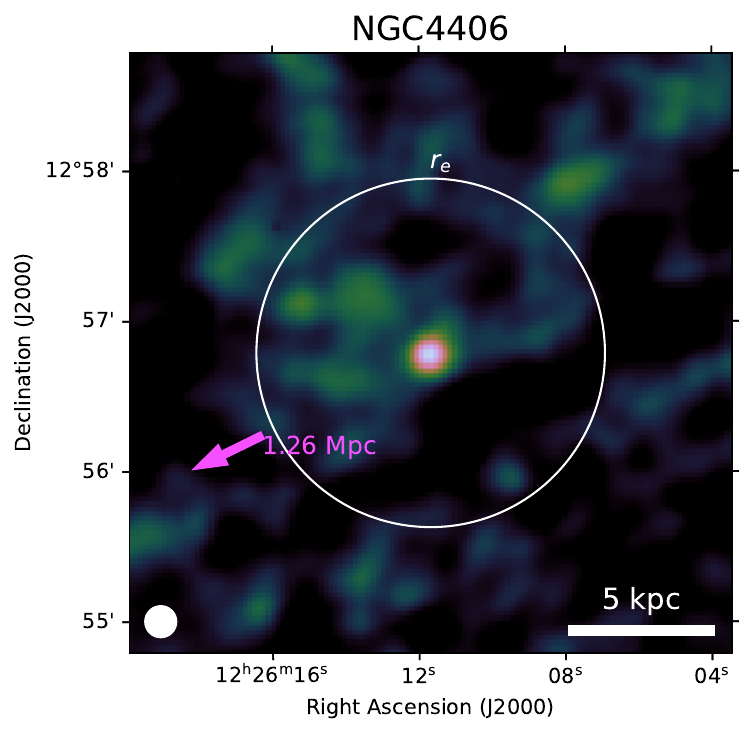}
         \label{fig:ngc4406_meerkat}
     \end{subfigure}
     \hfill
     \begin{subfigure}[h]{0.26\textwidth}
         \centering
         \includegraphics[width=\textwidth]{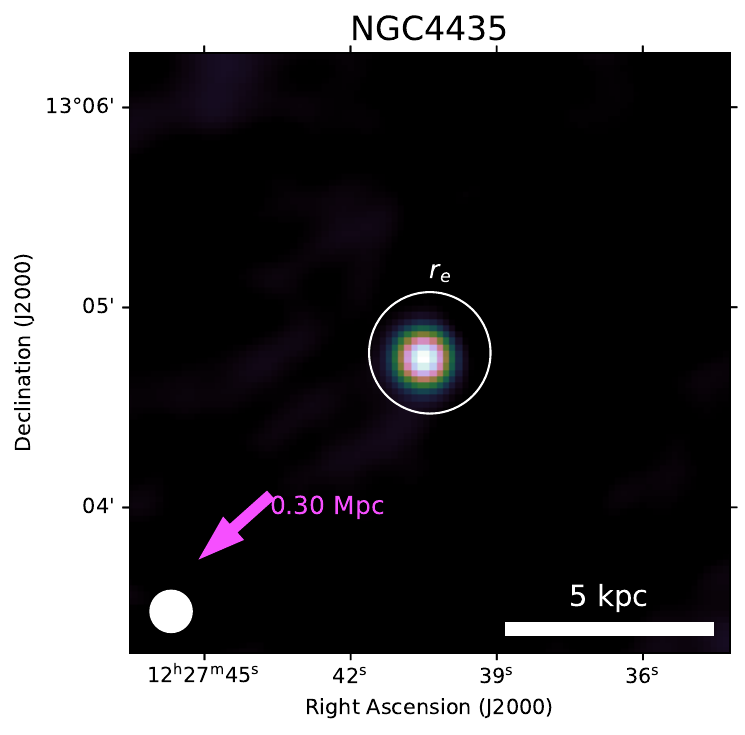}
         \label{fig:ngc4435_meerkat}
     \end{subfigure}
     \begin{subfigure}[h]{0.26\textwidth}
         \centering
         \includegraphics[width=\textwidth]{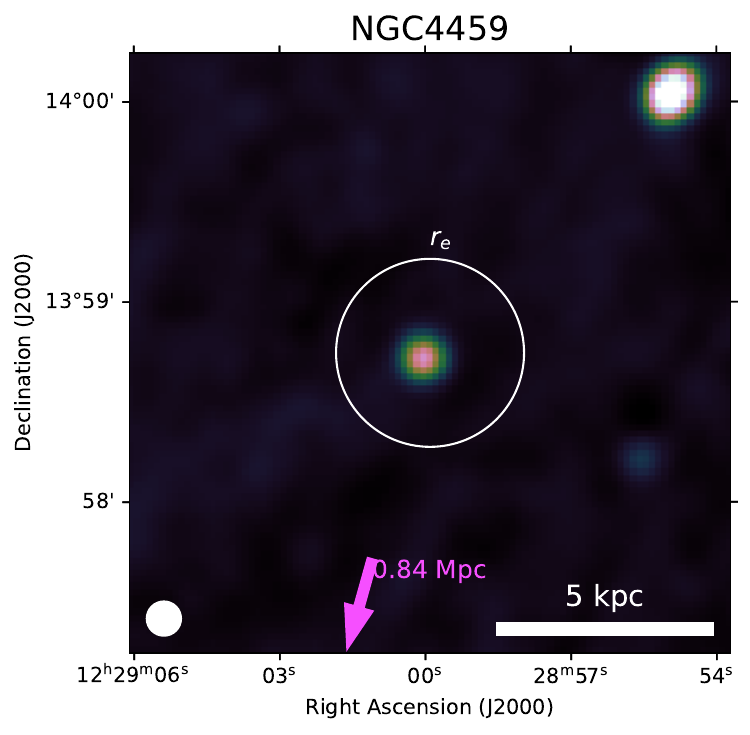}
         \label{fig:ngc4459_meerkat}
     \end{subfigure}
     \hfill
     \begin{subfigure}[h]{0.26\textwidth}
         \centering
         \includegraphics[width=\textwidth]{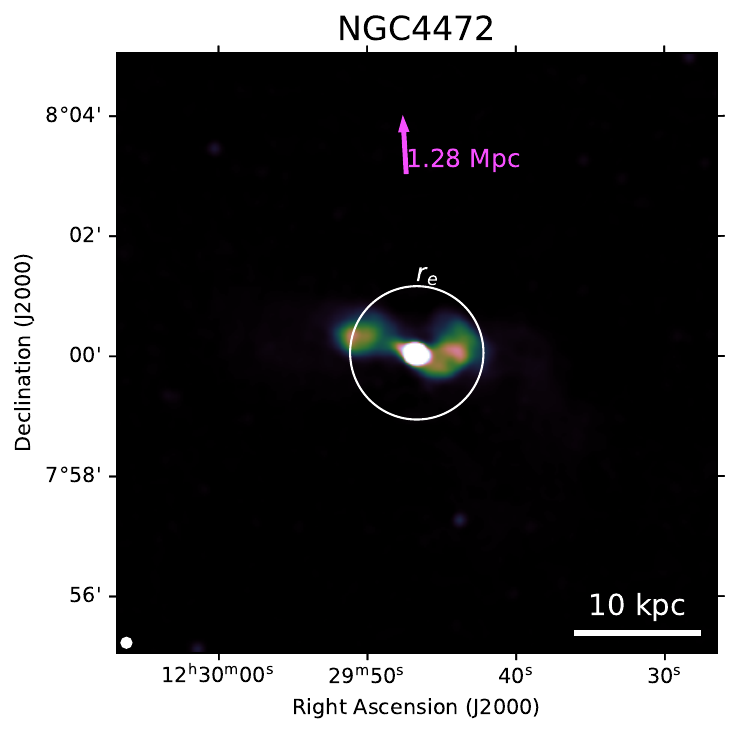}
         \label{fig:ngc4472_meerkat}
     \end{subfigure}
     \hfill
     \begin{subfigure}[h]{0.26\textwidth}
         \centering
         \includegraphics[width=\textwidth]{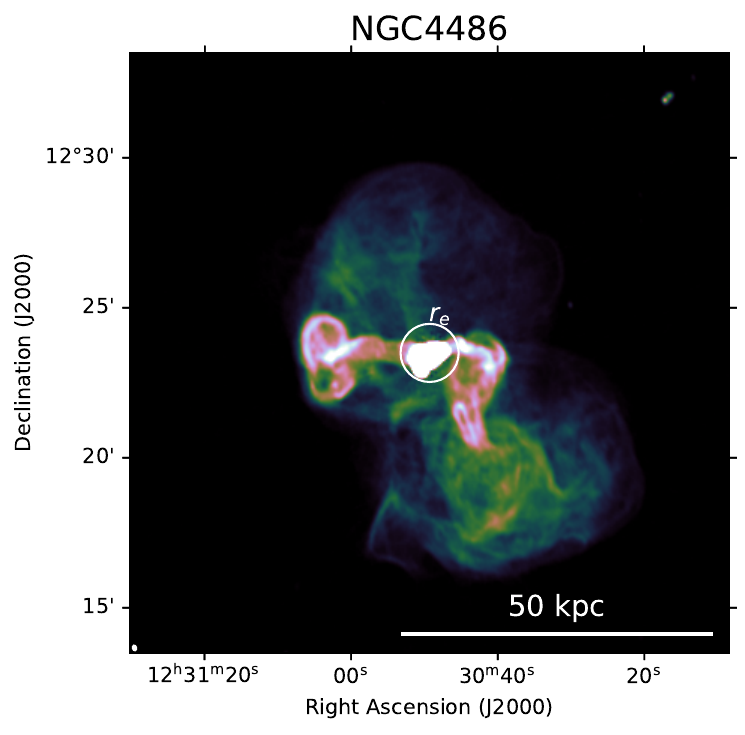}
         \label{fig:ngc4486_meerkat}
     \end{subfigure}
     \begin{subfigure}[h]{0.26\textwidth}
         \centering
         \includegraphics[width=\textwidth]{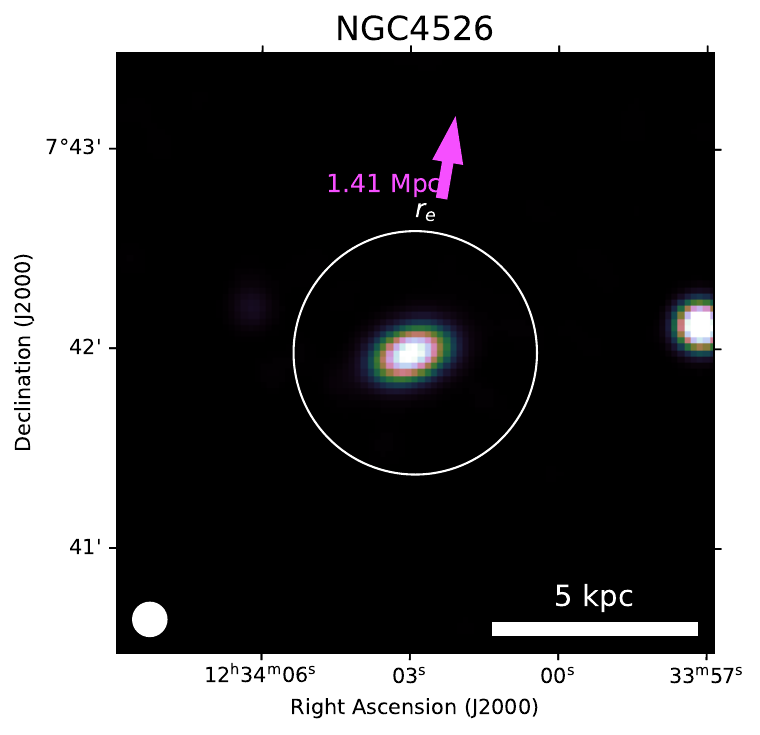}
         \label{fig:ngc4526_meerkat}
     \end{subfigure}
     \hfill
     \begin{subfigure}[h]{0.26\textwidth}
         \centering
         \includegraphics[width=\textwidth]{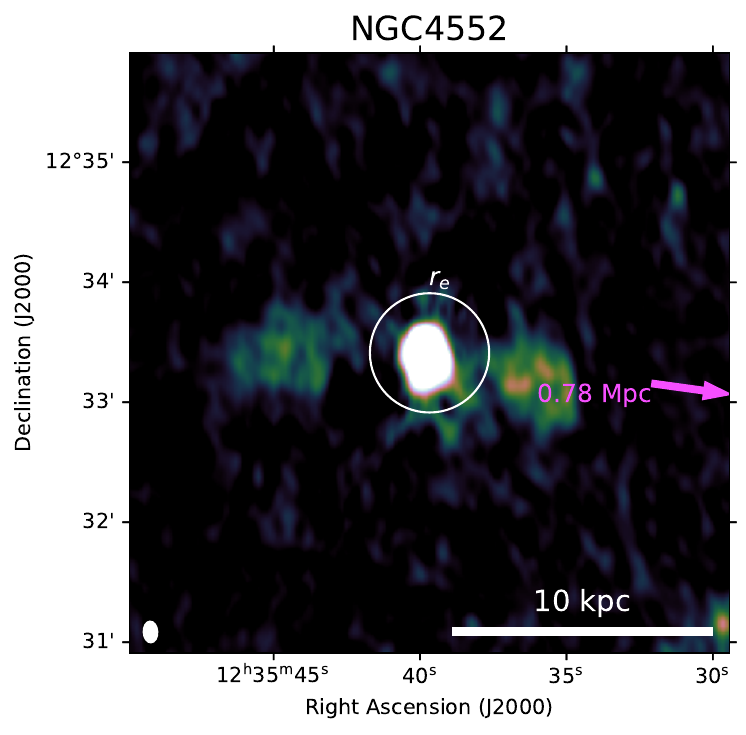}
         \label{fig:ngc4552_meerkat}
     \end{subfigure}
     \hfill
     \begin{subfigure}[h]{0.26\textwidth}
         \centering
         \includegraphics[width=\textwidth]{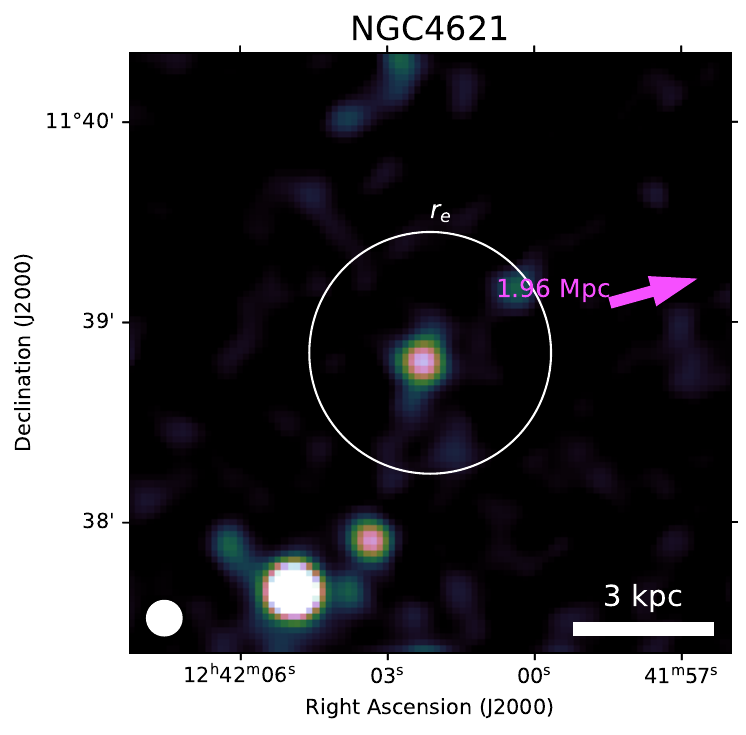}
         \label{fig:ngc4621_meerkat}
     \end{subfigure}
     \begin{subfigure}[hl]{0.26\textwidth}
         \centering
         \includegraphics[width=\textwidth]{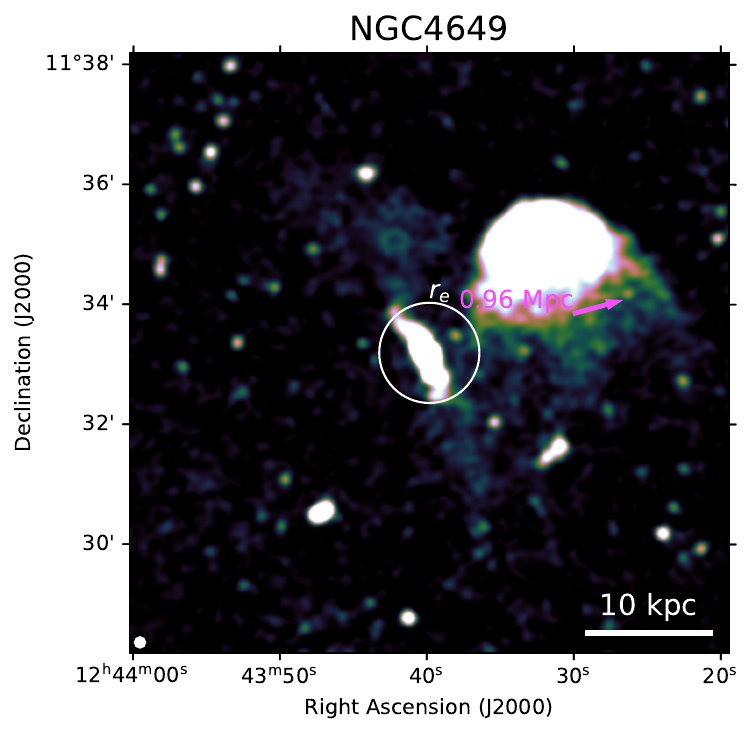}
         \label{fig:ngc4649_meerkat}
     \end{subfigure}
     \hfill
     \begin{subfigure}[hl]{0.26\textwidth}
         \centering
         \includegraphics[width=\textwidth]{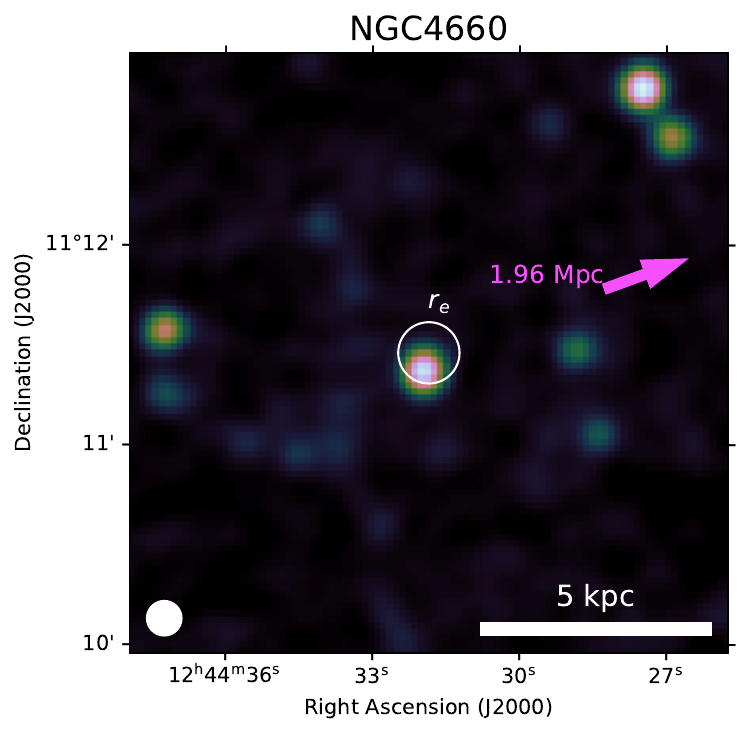}
         \label{fig:ngc4660_meerkat}
     \end{subfigure}
     \hfill
    \caption{MeerKAT image at 1.28\,GHz of 11 of the 12 sources from the sample covered by MeerKAT observations. As before each source is overlayed with the half light radius in the r-band of its host galaxy (white circle) taken from \cite{2014ApJS..215...22K}. The pink arrow in each image marks the direction and physical distance towards the centre of the Virgo cluster (NGC\,4486). The ellipse in the lower left corner indicates the beam size between $9.7''$ and $12.7''$}
    \label{fig:radio_overlays}
\end{figure*}

\section{Spectral index uncertainty maps}
\label{sec:app_spidxerr}
\begin{figure*}[h]
    \centering
    \begin{subfigure}[h]{0.3\columnwidth}
            \centering
            \includegraphics[width=\linewidth]{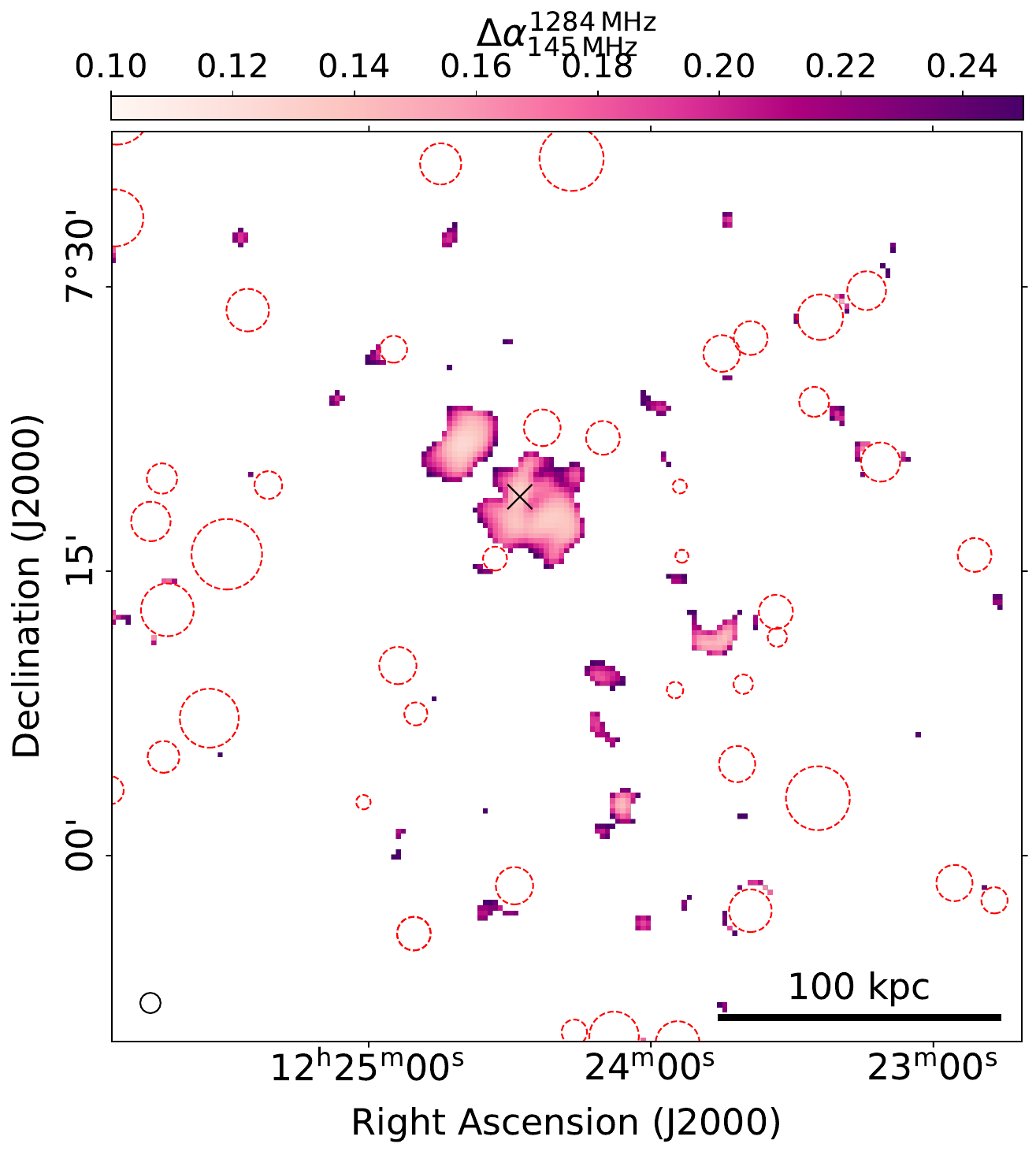}
        \caption{NGC\,4365}
            \label{fig:spidx_4365_err}
        \end{subfigure}
    \begin{subfigure}[h]{0.3\columnwidth}
        \centering
        \includegraphics[width=\linewidth]{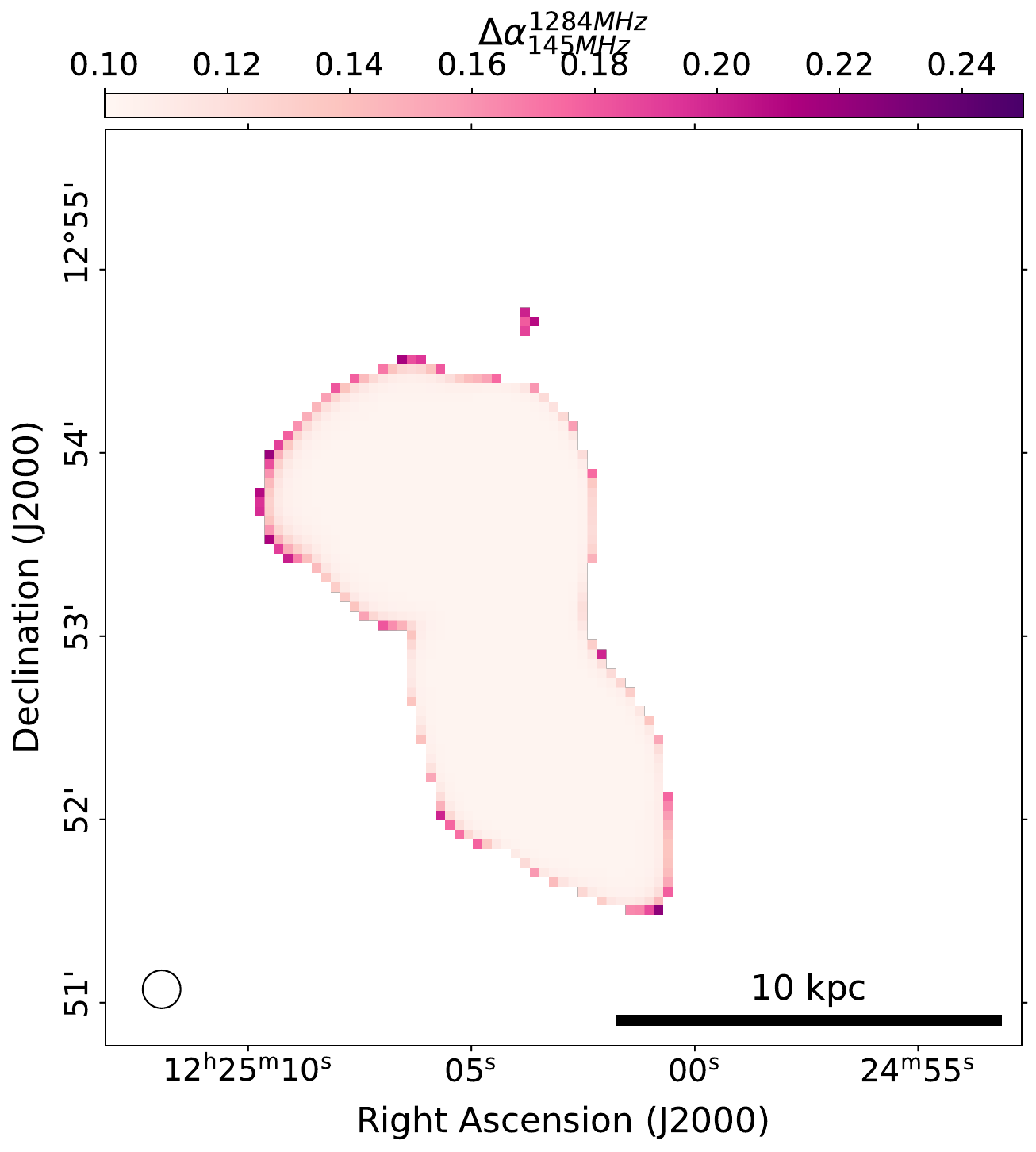}
        \caption{NGC\,4374}
        \label{fig:spidx_4374_err}
    \end{subfigure}
    \begin{subfigure}[h]{0.3\columnwidth}
        \centering
        \includegraphics[width=\linewidth]{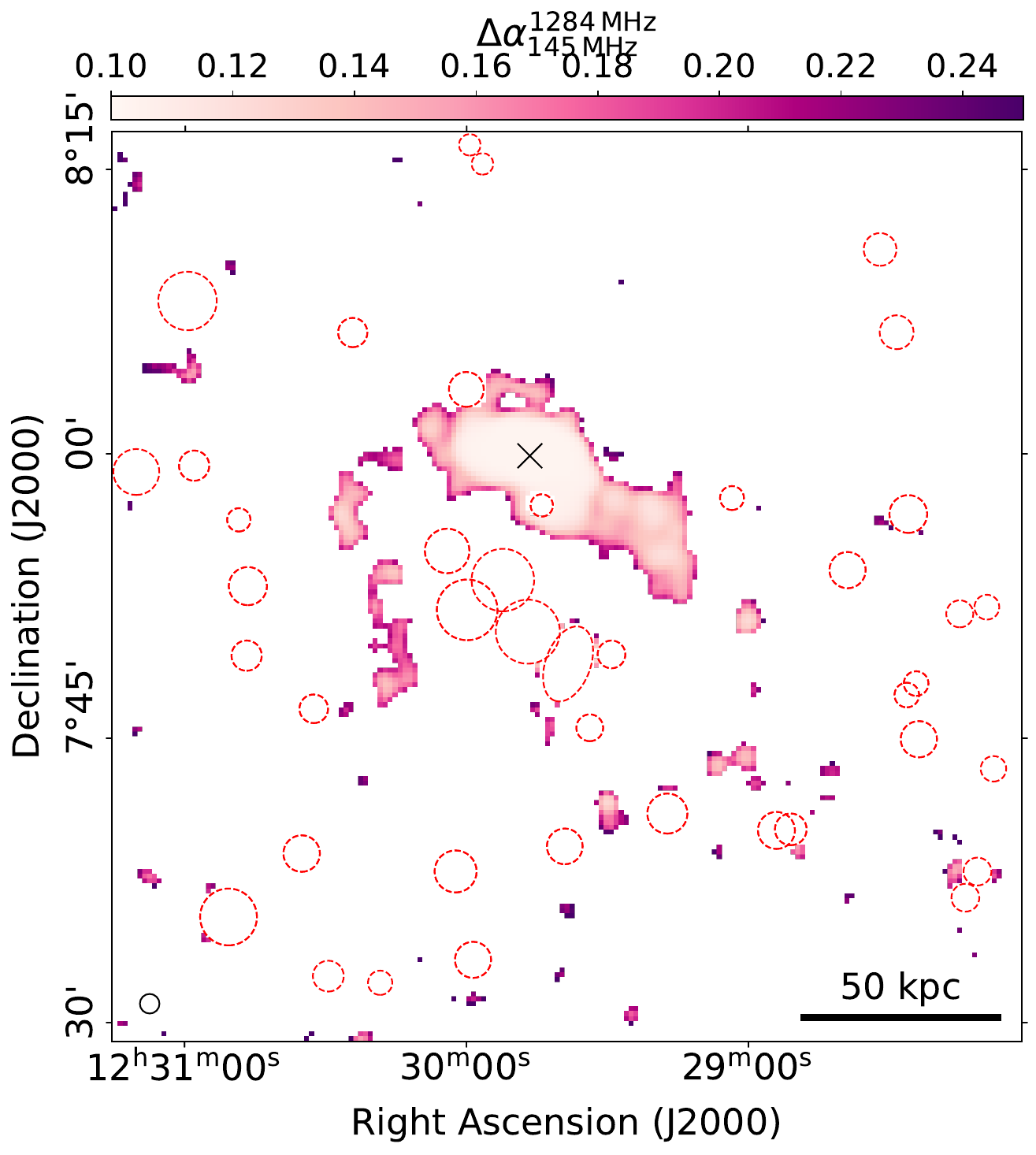}
        \caption{NGC\,4472}
        \label{fig:spidx_4472_err}
    \end{subfigure}
    \begin{subfigure}[h]{0.3\columnwidth}
        \centering
        \includegraphics[width=\linewidth]{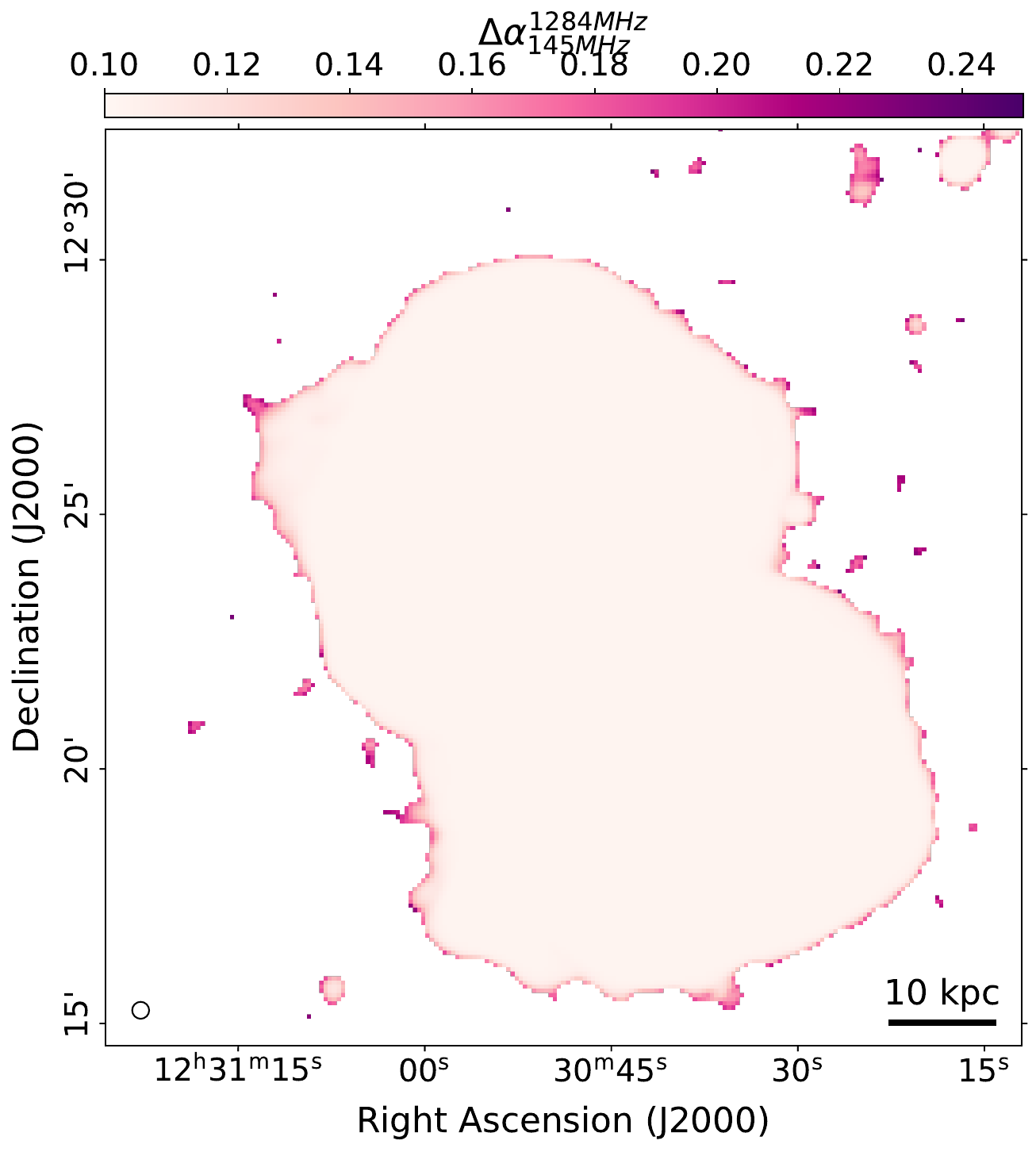}
        \caption{NGC\,4486}
        \label{fig:spidx_4486_err}
    \end{subfigure}
    \begin{subfigure}[h]{0.3\columnwidth}
            \centering
            \includegraphics[width=\linewidth]{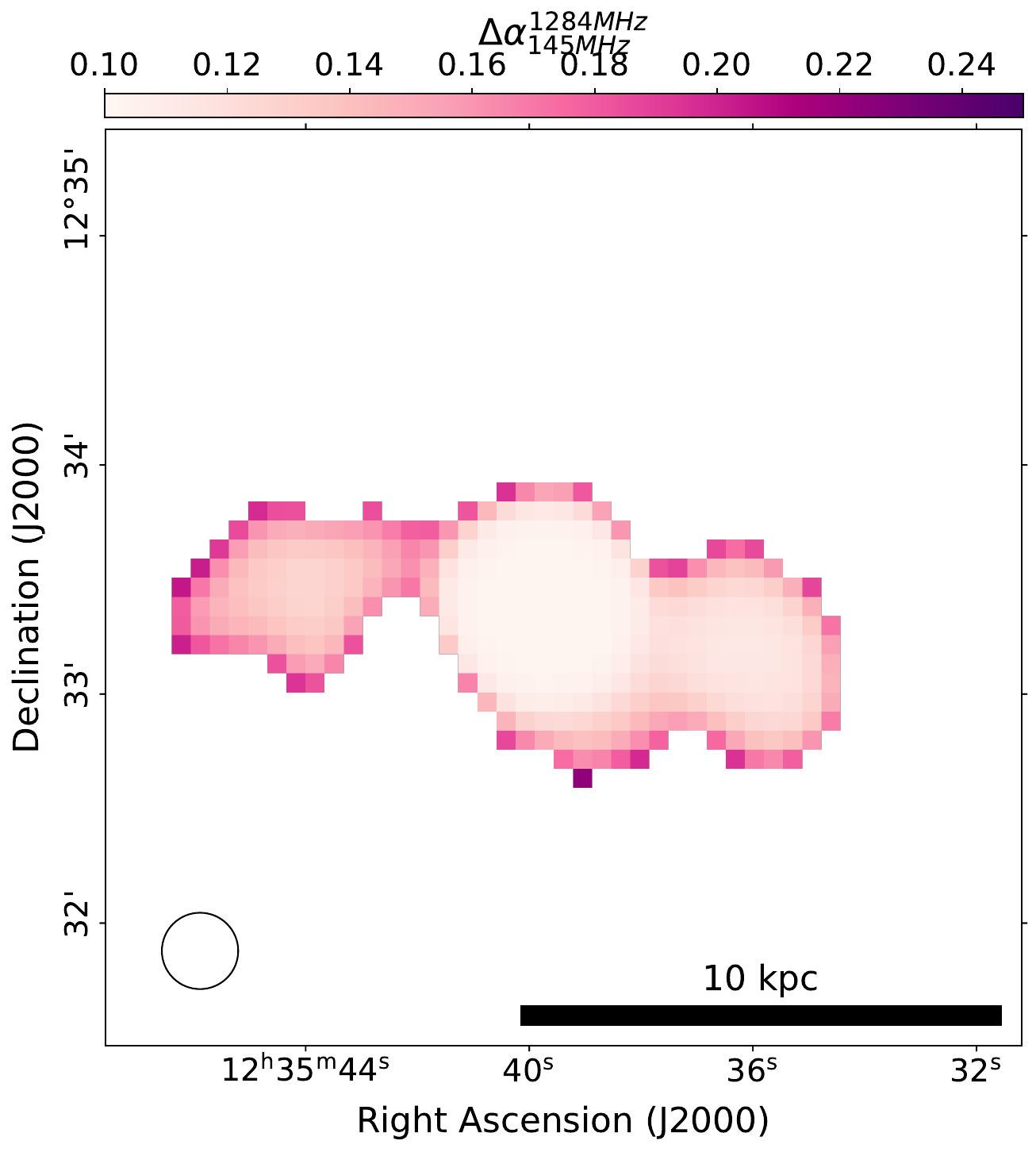}
        \caption{NGC\,4552}
            \label{fig:spidx_4552_err}
    \end{subfigure}
    \begin{subfigure}[h]{0.3\columnwidth}
            \centering
            \includegraphics[width=\linewidth]{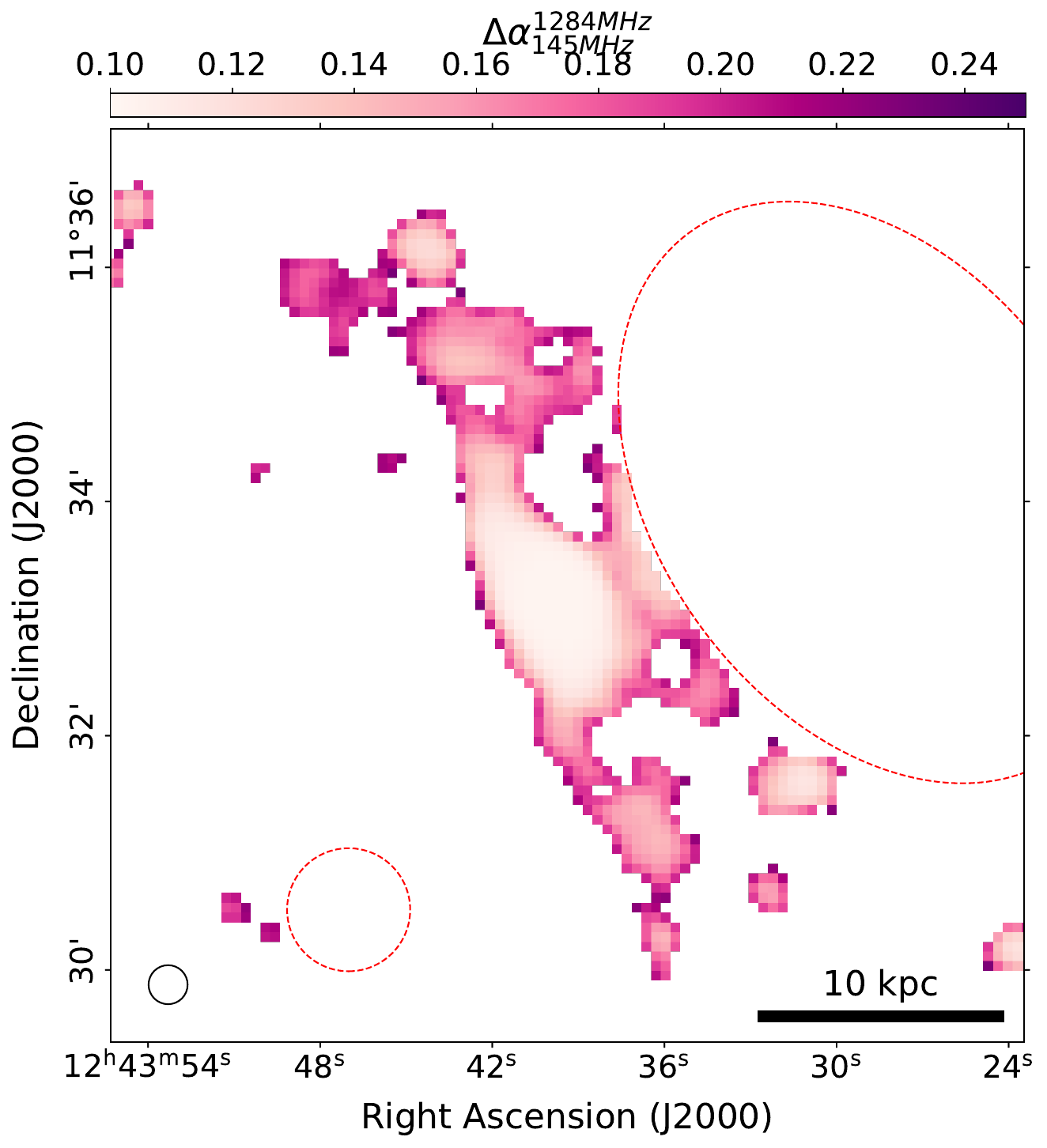}
        \caption{NGC\,4649}
            \label{fig:spidx_4649_err}
        \end{subfigure}
    \caption{Spectral index uncertainty maps of the six galaxies with extended radio emission. Dashed red circles mark unrelated sources that were masked. The uncertainties include the systematic uncertainty on the flux density scale for LOFAR at 20\% and for MeerKAT at 10\%.}
    \label{fig:spidx_maps_err}
\end{figure*}

\end{appendix}
\end{document}